\documentclass[manuscript=article]{achemso}

\usepackage{upgreek}
\usepackage[version=3]{mhchem} 
\usepackage[T1]{fontenc}       
\usepackage{graphicx}					
\usepackage{epstopdf}
\usepackage{amsmath}
\usepackage{color}

\usepackage{subcaption}
\usepackage{hyperref}
\usepackage[]{multirow}
\usepackage[]{multicol}
\usepackage[]{float}
\usepackage[]{rotating}
\usepackage{listings}
\setkeys{acs}{usetitle = true}

\title{Influence of Core Substitution on the Electronic Structure of Benzobisthiadiazoles}

\author{Mohsen Ajdari}
\affiliation{Physikalisch-Chemisches Institut, Universit\"at Heidelberg, Im Neuenheimer Feld 253, 69120 Heidelberg, Germany}
\author{Ronja Pappenberger}
\affiliation{Physikalisch-Chemisches Institut, Universit\"at Heidelberg, Im Neuenheimer Feld 253, 69120 Heidelberg, Germany}
\author{Caja Annweiler}
\affiliation{Physikalisch-Chemisches Institut, Universit\"at Heidelberg, Im Neuenheimer Feld 253, 69120 Heidelberg, Germany}
\author{Tobias Kaczun}
\affiliation{Interdisziplin\"{a}res Zentrum f\"{u}r Wissenschaftliches Rechnen, Universit\"at Heidelberg, Im Neuenheimer Feld 205A, 69120 Heidelberg, Germany}
\author{Leon M\"{u}ller}
\affiliation{Interdisziplin\"{a}res Zentrum f\"{u}r Wissenschaftliches Rechnen, Universit\"at Heidelberg, Im Neuenheimer Feld 205A, 69120 Heidelberg, Germany}
\author{Larissa Winkelmann}
\affiliation{Physikalisch-Chemisches Institut, Universit\"at Heidelberg, Im Neuenheimer Feld 253, 69120 Heidelberg, Germany}
\author{Lukas Ahrens}
\affiliation{Organisch-Chemisches Institut, Universit\"at Heidelberg, Im Neuenheimer Feld 270, 69120 Heidelberg, Germany}
\author{Uwe H. F. Bunz}
\affiliation{Organisch-Chemisches Institut, Universit\"at Heidelberg, Im Neuenheimer Feld 270, 69120 Heidelberg, Germany}
\author{Andreas Dreuw}
\affiliation{Interdisziplin\"{a}res Zentrum f\"{u}r Wissenschaftliches Rechnen, Universit\"at Heidelberg, Im Neuenheimer Feld 205A, 69120 Heidelberg, Germany}
\affiliation{Physikalisch-Chemisches Institut, Universit\"at Heidelberg, Im Neuenheimer Feld 253, 69120 Heidelberg, Germany}
\author{Petra Tegeder}
\affiliation{Physikalisch-Chemisches Institut, Universit\"at Heidelberg, Im Neuenheimer Feld 253, 69120 Heidelberg, Germany}
\email{tegeder@uni-heidelberg.de}
\phone{+49 (0) 6221 548475}

\date{\today}
\begin{document}
\begin{abstract}
Benzobisthiadiazoles (BBTs) are promising organic semiconductors for applications in field effect transistors and solar cells, since they possess a strong electron-accepting character. Thereby the electronic structure of organic/metal interfaces and within thin films is essential for the performance of organic electronic devices.
Here, we study the structural and the electronic properties of two BBTs, with different core substitution pattern, a phenyl (BBT-Ph) and thiophene (BBT-Th) derivative adsorbed on Au(111) using vibrational and electronic high-resolution electron energy loss spectroscopy in combination with state-of-the-art quantum chemical calculations. In the mono- and multilayer both BBTs adopt a planar adsorption geometry with the molecular backbone as well as the phenyl and thiophene side groups are oriented parallel to the gold substrate. The energies of the lowest excited electronic singlet states ($S$) and the first triplet state ($T_{1}$) are determined.
The optical gap ($S_{0} \rightarrow S_{1}$ transition) is found to be 2.2 eV for BBT-Ph and 1.6 eV for BBT-Th. The energy of $T_{1}$ is identified to be 1.2 eV in BBT-Ph and in the case of BBT-Th 0.7 eV. Thus, both the optical gap size as well as the $T_{1}$ energy are drastically reduced in BBT-Th compared to BBT-Ph. Based on our quantum chemical calculations this is attributed to the electron-rich nature of the five-membered thiophene rings in conjunction with their preference for planar geometries. Variation of the substitution pattern in BBTs opens the opportunity for tailoring their electronic properties.

\end{abstract}
\maketitle

\section{Introduction}
Benzo[1,2-c:4,5-c']bis[1,2,5]thiadiazole (BBT) containing polymers and small molecules are promising candidates for applications in organic (opto) electronic devices such as solar cells, light emitting diodes, or field effect transistors \cite{Tam2015, Yuen2011, Ye2019, Tam2012}, because they possess a strong electron-accepting character \cite{Muller2019} and inherent small optical band gaps \cite{Thomas2012, Chmovzh2018, Yamashita1997, Qian2010}. In addition, they show auspicious charge carrier mobilities, often ambipolar \cite{Yuen2011, Yuen2011a}. For device performance the molecular orientation at the interface to an electrode as well as within a molecular film (film morphology) plays an essential role, since it strongly influences the electronic properties such as energy level alignment or charge transport properties \cite{Ishii1999,Koch2013,Gruenewald2013,Braun2009,Oehzelt2014,Mercurio2013, Bredas2009,Koehler2015,May2011,Beljonne2011,Ruhle2011,Avino2016,Casu2015,Forker2012}.
Utilizing
high-resolution electron energy loss spectroscopy (HREELS) allows for the analysis of both the adsorption and electronic properties of molecules on (semi)metallic  surfaces and within thin films \cite{Bronner2012, Bronner2013a, Maass2017, Maass2016, maass2015, Stein2017, Stremlau2017, Gahl2013, Leyssner2010, Ovari2007, Navarro2014, Ajdari2020}. HREELS has already been successfully applied to investigate the energies of intramolecular electronic transitions (e.g. $S_{0} \rightarrow S_{1}$) \cite{Bronner2012, Bronner2013a, Maass2017, Maass2016, maass2015, Navarro2014, Ajdari2021, Ajdari2023}.
Furhermore, HREELS opens up the opportunity to gain insights into the energetic position of triplet states \cite{Swiderek1990, Swiderek1994, Ajdari2020b, Hoffmann2022, Ajdari2021}, which are not accessible with (linear) optical methods. While the adsorption and electronic properties of S-heteropolycyclic aromatic molecules in particular thiophene derivatives on metal substrates have been studied in great detail (see e.g. Refs. \cite{Hill2000, Koch2005, Kiguchi2004,Yokoyama2006, Kakudate2006, Kiel2007, Grobosch2007, Varene2011, Varene2012, Varene2012a, bogner2015electronic, bogner2016electronic, Bronsch2019}) due to their relevance in vacuum-processed small molecule organic solar cells \cite{Mishra2011, Ziehlke2011, Fitzner2011, Fitzner2012,Meerheim2014}, only a few examples of N-heteropolycycles are known in literature \cite{maass2015, Maass2016, Ajdari2020, Ajdari2020b, Stein2021, Stein2021b,  Hoffmann2022, Ajdari2023}. Notably, N-heteropolycyclic compounds are promising candidates for electron transporting (n-channel) semiconductors, which are of great interest as organic field effect transistors \cite{klauk2007, wurthner2011, Bunz2013, miao2014ten, bunz2015larger}. In the case of S- and N-containing heteropolycycles only one study is known in literature \cite{Ajdari2021}. Therein, the structural and electronic properties of naphthothiadiazole derivatives adsorbed on Au(111) have been investigated with HREELS, in particular the influence of core halogenation on the electronic structure \cite{Ajdari2021}. Several singlet and the first triplet transition energies have been determined. It has been found that halogenation leads to a decrease of the optical gap size.
\begin{figure}[htb]
\centering
\resizebox{0.4\hsize}{!}{\includegraphics{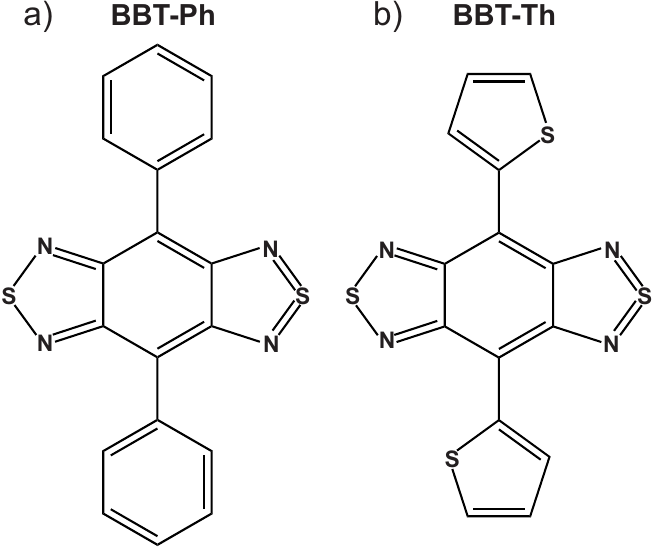}}
\caption{Benzo[1,2-c:4,5-c']bis[1,2,5]thiadiazole (BBT) derivatives investigated in the present study.}
\label{molecules}
\end{figure}

In the present contribution we study the adsorption and electronic properties of two BBT derivatives, the phenyl substituted 4,8-diphenyl-benzo[1,2-c:4,5-c']bis[1,2,5]thiadiazole (BBT-Ph) and
 thiophene substituted 4,8-dithiophene-benzo[1,2-c:4,5-c']bis[1,2,5]thiadiazole (BBT-Th)
shown in Fig. \ref{molecules} on Au(111) using vibrational and electronic HREELS as well as state-of-the-art quantum chemical methodology.
In particular we focus on the influence of core substitution (phenyl vs. thiophene) on the electronic structure. We found that both BBT derivatives in the monolayer and thin film adopt an adsorption geometry in which the molecular backbone and the side groups (phenyl or thiophene) are orient themselves parallel to the Au(111) substrate. For both derivatives several singlet transitions are identified as well as the first triplet state. Going from BBT-Ph to BBT-Th results in a pronounced reduction of the electronic transition energies.

\section{Methods}
The HREELS experiments were performed under ultra-high vacuum conditions at a sample temperature of around 90 K.
The Au(111) single crystal was prepared by standard procedure of Ar$^{+}$ sputtering and annealing.
The BBT compounds were synthesized according to the procedures reported in Ref. \cite{Li2011}. They were deposited from an effusion cell held at 443 K for BBT-Ph and at 468 K for BBT-Th onto the Au(111) sample held at 300 K.
The coverage was determined by temperature-programmed desorption (TPD) measurements (see supporting information; Figs. S1 and S2)). While in the monolayer (ML) regime i.e., the molecules in direct contact with the metal substrate,  BBT-Ph desorbs intact, the BBT-Th molecules are decaying as can be concluded from HREELS and scanning tunneling microscopy experiments (see supporting information; Figs. S3--S5), therefore the coverage determination for BBT-Th has a larger error. However, for both BBT-Ph a defined ML coverage can be prepared by adsorbing a multilayer followed by heating to 400 K to desorb the multilayer. For BBT-Th direct deposition of the molecules was used.
HREELS measurements were performed with incident electron energy of 3.5 eV and 15 eV for vibrational and electronic HREELS, respectively.
For experimental details see Refs. \cite{ibach1982, Tegeder2012, maass2015, Maass2016, Maass2017}.
To assign the vibrational modes of BBT-Ph and BBT-Th, density functional theory (DFT) calculations for the isolated molecules were carried out using the Gaussian09 package \cite{gaussian2009}.
Thereby the standard B3LYP exchange-correlation (xc) functional and the 6-311G basis set were employed.

Initial input geometries were generated using the universal force field as implemented in Avogadro\cite{hanwell:2012:AvogadroAdvancedSemantic}
and then optimized using DFT \cite{hohenberg:1964:InhomogeneousElectronGasa, kohn:1965:SelfConsistentEquationsIncluding} with the standard B3LYP\cite{becke:1993:DensityFunctionalThermochemistry} xc functional and the aug-cc-pVDZ\cite{woon:1993:GaussianBasisSets, kendall:1992:ElectronAffinitiesFirst, dunning:1989:GaussianBasisSets} basis set employing the D4 dispersion correction\cite{caldeweyherGenerallyApplicableAtomicCharge2019}.
Frequency calculations were performed to confirm the nature of the stationary points as minima on the potential energy surface.
The conductor-like polarizable continuum model (CPCM) was utilized to account for solvent
effects of dichloromethane (DCM, $\epsilon=9.08$) and chloroform (\ce{CHCl3}, $\epsilon = 4.9$).\cite{neese:2018:SoftwareUpdateORCA, barone:1998:QuantumCalculationMolecular} 
Excitation energies were calculated at the optimized ground state equilibrium geometries with the Tamm-Dancoff approximation (TDA)\cite{hirata:1999:TimedependentDensityFunctional} to linear-response time-dependent DFT (TDDFT)\cite{dreuw:2005:SingleReferenceInitioMethods, herbert:2023:DensityFunctionalTheory} employing the BMK\cite{boese:2004:DevelopmentDensityFunctionals} xc-functional and aug-cc-pVDZ basis set.

\section{Results and Discussion}
While the energy level alignment at metal/organic interfaces strongly depends on the electronic coupling between metal and molecular states, the electronic properties of the molecules in a film are strongly influenced by the strength of their intermolecular interactions.
In addition, electronic properties are also highly affected by the molecular adsorption geometry, both at the interface as well as within the film.
Therefore we first analyzed the adsorption properties of the BBT derivatives at the interface and within the thin films using vibrational HREELS.
\subsection{Adsorption properties of benzobisthiadiazoles on Au(111)}
\begin{figure}[h!]
\centering
\resizebox{0.45\hsize}{!}{\includegraphics{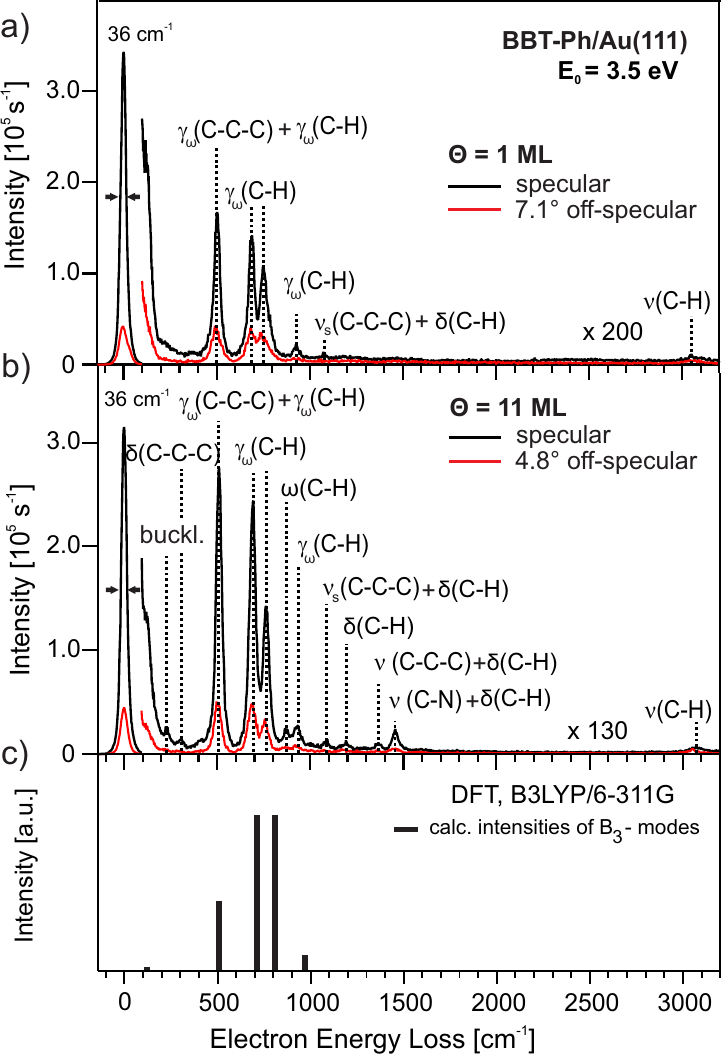}}
\caption{Vibrational HREEL spectra recorded in specular (black) and off-specular (red) scattering geometry of a) 1 ML and b) 11 ML BBT-Ph adsorbed on Au(111). $E_{0}$ is the primary energy of the incident electrons. The energy resolution of 36 cm$^{-1}$ is determined by the full width at half maximum (FWHM) of the elastic peak  (zero energy loss peak). c) DFT-calculated intensities and frequencies of B$_{3}$-symmetric vibrations which possess transition dipole moments perpendicular to the molecular backbone.}
\label{vib_BBT-Ph}
\end{figure}

In Figure \ref{vib_BBT-Ph} a) the HREEL spectra of 1 ML  BBT-Ph measured in specular and 7.1$^{\circ}$ off-specular scattering geometry are displayed, while the respective data for a coverage of 11 ML (off-specular data have been measured at 4.8$^{\circ}$) are presented in Figure \ref{vib_BBT-Ph} b).
In addition, the calculated intensities and frequencies of vibrational modes possessing a transition dipole moment perpendicular to the molecular backbone are shown in Figure \ref{vib_BBT-Ph} c). These modes belong to the irreducible representation $B_{3}$  of the $D_{2}$ molecular point group of BBT-Ph.
In the spectra of 1 ML and 11 ML three pronounced vibrations located around  499, 690, and 748 cm$^{-1}$ (values obtained for 1 ML BBT-Ph/Au(111)) are observed. These vibrational modes are dipole-active,  i.e., the intensity ratio between specular and off-specular scattered electrons is high and dipole scattering is the main excitation mechanism.
By comparing
the vibrational HREEL spectrum with the DFT calculation, the vibrations can
be assigned to a combined
$\gamma_{\omega}$(C-C-C) and
$\gamma_{\omega}$(C-H) wagging ($\omega$) mode in the phenyl moieties  (499 cm$^{-1}$) and to
$\gamma_{\omega}$(C-H) wagging vibrations in the phenyl side groups (690 and 748 cm$^{-1}$) (for the assignment see Table \ref{vibrations_BBT-Ph} and the supporting information for the visualization of the vibrational modes, Fig. S6).

\begin{table}[h]
\centering
\caption{Vibrational modes in cm$^{-1}$ of BBT-Ph for 1 ML and 11 ML adsorbed on Au(111).  The abbreviation \textit{da} denotes dipole-active modes. In addition, DFT calculated frequencies based on the B3LYP functional and the 6-311G basis set of the gas phase molecules are displayed. Further abbreviations: $\nu$ -- stretching; $\delta$ -- deformation; $\omega$ -- wagging; $\tau$ -- twisting; $\gamma$ -- out-of-plane; Repr. -- representation of the point group and in brackets corresponding orientation of the calculated dipole derivative vector with respect to the molecular geometry, y BBT-backbone axis, z residual group axis, x perpendicular to the BBT-Ph plane.}
\begin{tabular}{cccccc}
\ \textbf{1 ML}  & \textbf{11 ML} & \textbf{DFT}& \textbf{Mode} & \textbf{Repr.}\\
\hline

- & $223$ & $242$ & buckling & $B_2(y)$\\
 - & $307$ & $294$ & $\delta$(C-C-C) & $B_2(y)$\\
  $499 \; \textit{da}$ & $510 \; \textit{da}$ & $507$ & $\gamma_{\omega}$(C-C-C), $\gamma_{\omega}$(C-H) & $B_3(x)$\\
  $690 \; \textit{da}$ & $694 \; \textit{da}$ & $711$ & $\gamma_{\omega}$(C-H) & $B_3(x)$\\
  $748 \; \textit{da}$ & $762 \; \textit{da}$ & $809$ & $\gamma_{\omega}$(C-H) & $B_3(x)$\\
 - & $866$ & $865$ & $\omega$(C-H) & $B_1(z)$\\
  $926 \; \textit{da}$ & $933 \; \textit{da}$ & $953$ & $\gamma_{\omega}$(C-H) & $B_3(x)$\\
  $1071$ & $1092$ & $1058$ & $\nu_{s}$(C-C-C), $\delta$(C-H)& $B_1(z)$\\
  - & $1192$ & $1214$ & $\delta$(C-H) & $B_2(y)$\\
  - & $1366$ & $1367$ & $\nu$(C-C-C), $\delta$(C-H) & $B_2(y)$\\
 - & $1456$ & $1459$ & $\nu$(CN), $\delta$(C-H) & $B_1(z)$\\
 $3044$ & $3071$ & $3165$ & $\nu$(C-H) & $B_1(z)$\\
\hline
\end{tabular}
\label{vibrations_BBT-Ph}
\end{table}

Another dipole active out-of-plane vibration is located at 926 cm$^{-1}$
which is associated with a
$\gamma_{\omega}$(C-H) wagging mode in the
phenyl side groups.
Contrary, the intensity of the in-plane $\nu$(C--H) stretching mode (3044 cm$^{-1}$) shows no difference between dipole and impact scattered electrons. Thus, this mode is non-dipole active and accordingly impact scattering is the main excitation mechanism. Only one further in-plane vibration resulting from $\nu$(C--C--C) stretching vibration or $\delta$(C--H) deformation mode are detected at 1071 cm$^{-1}$, which possesses a very low intensity.  From these observations we conclude that in the monolayer regime BBT-Ph molecules adopt a planer adsorption geometry with the molecular backbone as well as the phenyl-rings oriented parallel to the Au(111) surface.
It has be mentioned,  that in the DFT calculations (after energy relaxation
and geometry optimisation) the molecule is not arranged planar. The
phenyl rings are twisted by 36.9$^{\circ}$ with respect to the planar molecular backbone. In comparison, in the BBT-Ph crystal structure an angle of 45.9$^{\circ}$ has been observed \cite{Yamashita1997}. The planar adsorption geometry results in  a change of the molecular point group from $D_{2}$ to
$D_{2h}$, which might be reason for the discrepancies between the vibrational transition energies attained from the DFT calculations in comparison to the experimental values (see Table \ref{vibrations_BBT-Ph}).
In the multilayer regime (Figure \ref{vib_BBT-Ph} b)), in addition to the very intense dipole-active modes found in the monolayer, several very low-intensity non-dipole active vibrations (below 510 cm$^{-1}$ and above 760 cm$^{-1}$) are observed, which exhibit a dynamic dipole moment parallel to the surface indicating a
slight tilting (or small disorder) of the molecular arrangement
at higher coverage.

\begin{figure}[h!]
\centering
\resizebox{0.45\hsize}{!}{\includegraphics{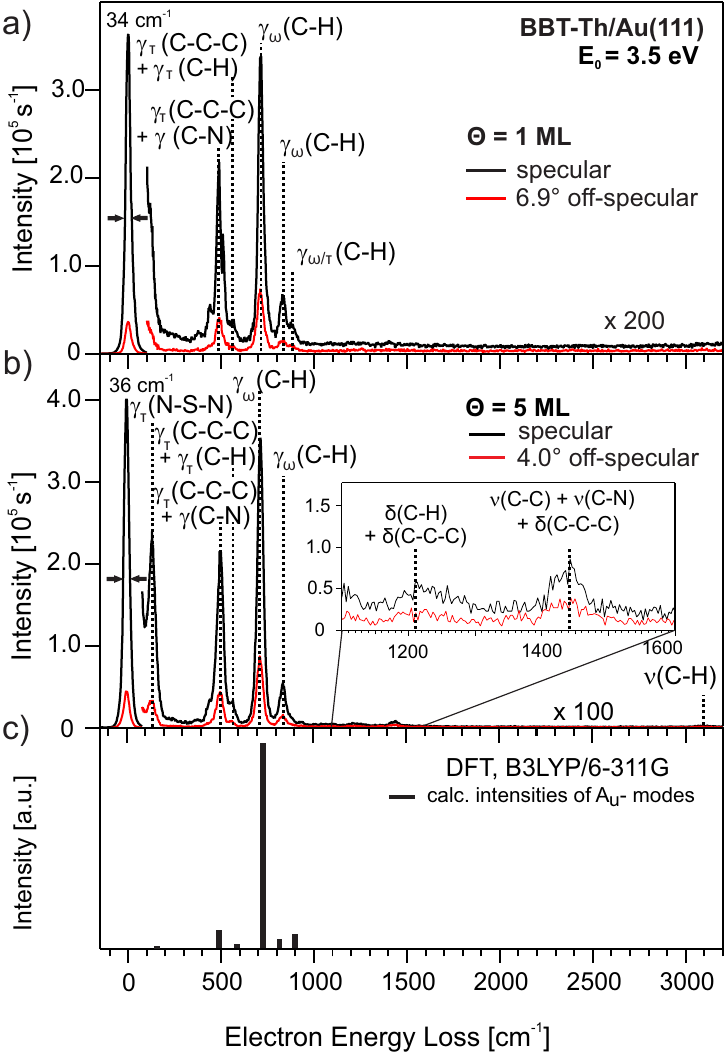}}
\caption{Vibrational HREEL spectra recorded in specular (black) and off-specular (red) scattering geometry of a) 1 ML and b) 5 ML BBT-Th/Au(111). $E_{0}$ is the primary energy of the incident electrons. The energy resolution of 34 cm$^{-1}$ (36 cm$^{-1}$) is determined by the full width at half maximum (FWHM) of the elastic peak  (zero energy loss peak). c) DFT-calculated intensities and frequencies of A$_{u}$-symmetric vibrations which possess transition dipole moments perpendicular to the molecular backbone.}
\label{vib_BBT-Th}
\end{figure}
The angle-resolved vibrational HREEL data for the BBT-Th in the mono- and multilayer regime as well as the calculated intensities and frequencies of vibrational modes possessing a dynamic dipole moment perpendicular to the molecular plane are displayed in Figure \ref{vib_BBT-Th}. In both the mono- and multilayer two very pronounced dipole active modes located at 490 and 713 cm$^{-1}$ for 1 ML BBT-Th/Au(111) can be assigned to the $\gamma_{\tau}$(C--C--C) out-of-plane twisting ($\tau$) of the thiophene substituent as well as  $\gamma$(C--N) mode (490 cm$^{-1}$)  and the $\gamma_{\omega}$(C--H) out-of-plane wagging mode of the thiophene moiety (713 cm$^{-1}$) (see Table \ref{vibrations_BBT-Th} for the assignment and the supporting information for the visualization of the vibrational modes, Fig. S7).

\begin{table}[h]
\centering
\caption{Vibrational modes in  cm$^{-1}$ of BBT-Th for 1 ML and 5 ML adsorbed on Au(111). The abbreviation \textit{da} denotes dipole-active modes. In addition DFT calculated frequencies based on the B3LYP functional and the 6-311G basis set of the free molecules are displayed. Further abbreviations: $\nu$ -- stretching; $\delta$ -- deformation; $\omega$ -- wagging; $\tau$ -- twisting; $\gamma$ -- out-of-plane; Repr. -- representation of the point group and in brackets corresponding orientation of the calculated dipole derivative vector with respect to the molecular geometry, x BBT-backbone axis, y residual group axis, z perpendicular to the BBT-Th plane.}
\begin{tabular}{ccccc}
\  \textbf{1 ML}  & \textbf{5 ML} & \textbf{DFT}& \textbf{Mode} & \textbf{Repr.}\\
\hline
 - & $137 \; \textit{da}$ & $161 $ & $\gamma_{\tau}$(N-S-N) & $A_u(z)$\\
$490 \; \textit{da}$ &$500 \; \textit{da}$ & $488$ & $\gamma_{\tau}$(C-C-C), $\gamma$(CN) & $A_u(z)$\\
 $567 \; \textit{da}$ & $571 \; \textit{da}$ & $585$ & $\gamma_{\tau}$(C-C-C), $\gamma_{\tau}$(C-H) & $A_u(z)$\\
 $713 \; \textit{da}$ & $719 \; \textit{da}$ & $724$ & $\gamma_{\omega}$(C-H) & $A_u(z)$\\
 $829 \; \textit{da}$ & $839 \; \textit{da}$ & $815$ & $\gamma_{\omega}$(C-H) & $A_u(z)$\\
 $874 \; \textit{da}$ & - & $899$ & $\gamma{\omega/\tau}$(C-H) & $A_u(z)$\\
 - & $1213$ & $1276$ & $\delta$(C-H), $\delta$(C-C-C) & $B_u(x,y)$\\
- & $1443$ & $1493$ & $\nu$(C-C), $\nu$(CN), $\delta$(C-C-C) & $B_u(x,y)$\\
 - & $3089$ & $3204$ & $\nu$(C-H) & $B_u(x,y)$\\
\hline
\end{tabular}
\label{vibrations_BBT-Th}
\end{table}
In the monolayer regime all observed vibrational losses are dipole-active, thus, as for the BBT-Ph, the BBT-Th molecules are adsorbed flat with the molecular backbone and the thiophene rings oriented parallel to the surface. Similar to BBT-Ph, further peaks with very low intensity appear in the mulitlayer spectrum above 1000 cm$^{-1}$ (Figure \ref{vib_BBT-Th}b)). They can be assigned to
in-plane modes. The most striking difference between the ML and multilayer data is the appearance of a very intense dipole-active peak at 137 cm$^{-1}$, which is attributed to the out-of-plane backbone
twisting mode $\gamma_{\tau}$(N--S--N). This can be explained by a strong adsorbate/substrate interaction in the monolayer regime which may lead to a suppression of this vibration. Hence, decoupling of the molecules from the metallic substrate at higher coverages results in the observation of the vibrational mode \cite{Kato2002}. However, we conclude that in the monolayer as well as in the multilayer regime the molecules adopt a planar adsorption geometry (molecular backbone and the thiophene moieties) with respect to the surface plane.

\subsection{Electronic properties of benzobisthiadiazoles on Au(111)}
For gaining comprehensive insights into the electronic structure of the BBTs, i.e., the identification and assignment of singlet and triplet transition energies and to analyze the influence of the different core substitution (phenyl vs. thiophene)  we utilized electronic HREELS and state-of-the-art quantum chemical methods.

\begin{figure}[htb]
\centering
\resizebox{0.45\hsize}{!}{\includegraphics{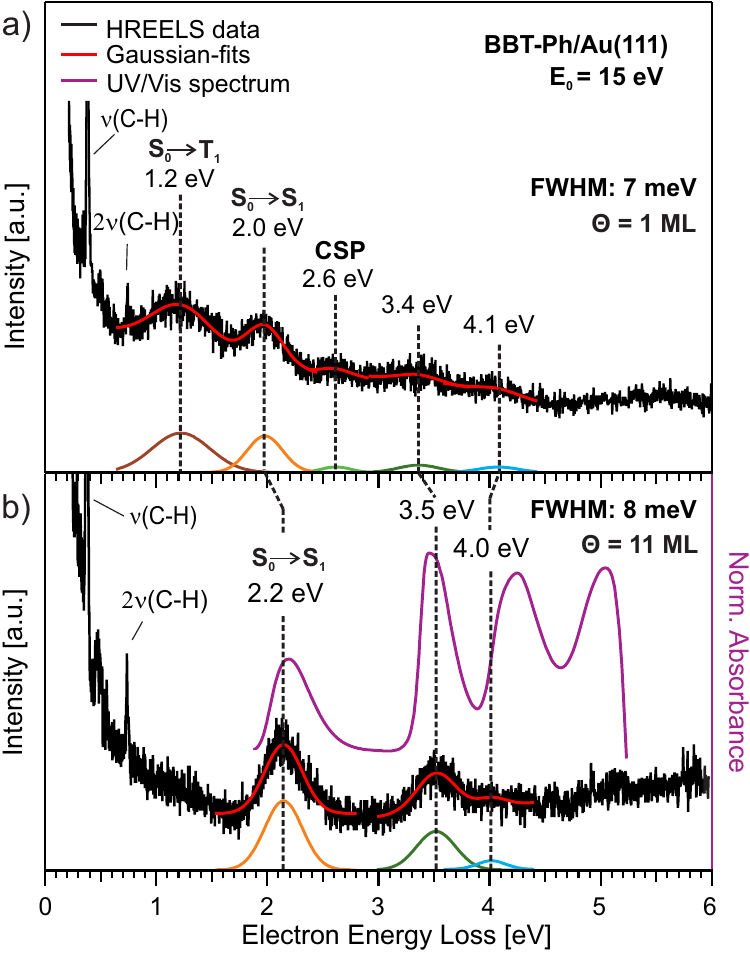}}
\caption{Electronic HREEL spectra of a)  1 ML and b) 11 ML BBT-Ph/Au(111) measured with an incident electron energy of 15 eV and under specular-scattering geometry. The electronic transitions were fitted using Gaussian functions (red curves). The BBT-Ph UV/vis spectrum obtained in hexane is shown for comparison (purple curve).}
\label{ele_BBT-Ph}
\end{figure}

Figure \ref{ele_BBT-Ph} shows  HREEL spectra of 1 ML and 11 ML BBT-Ph/Au(111) recorded with a primary electron energy of 15 eV.
In the 1 ML BBT-Ph/Au(111) spectrum several transitions are detected. They are located at 1.2, 2.0, 2.6, 3.4, and 4.1 eV (peak maximum).
At a coverage of 11 ML some features are more pronounced. In the ML spectrum we assign based on our quantum chemical calculations (see below) the peak detected at 1.2 eV to  the first triplet state ($T_{1}$) of BBT-Ph.
The peak at 2.0 eV is attributed to the $S_{0} \rightarrow S_{1}$ transition, thus to the optical gap in accordance with literature \cite{Qian2010}. The feature at 2.6 eV is connected with the conventional surface plasmon (CSP) of the Au(111) surface \cite{Park2009, Park2010}. The weak contributions at 3.4 and 4.1 eV are ascribed to molecular transitions. The former one gains intensity in the 11 ML spectrum as well as the $S_{0} \rightarrow S_{1}$ transition, because in the multilayer regime the molecules are electronically decoupled from the metallic
 substrate \cite{Maass2017, Maass2016, maass2015, Ajdari2020, Ajdari2020b,Ajdari2021}. The decoupling also leads to an increase of the transition energies, for instance the optical gap size rises by 200 meV from 2.0 to 2.2 eV. The plasmonic contribution (CSP) is not observed in the multilayer data due to the surface sensitivity of HREELS, viz. only the uppermost layers are probed.
 On the basis of our quantum chemical calculations (see below) we assign the peak at 3.5 eV and 4.0 eV to the $S_{0} \rightarrow S_{7}$ and $S_{0} \rightarrow S_{10}$ transition, respectively.

\begin{figure}[htb]
\centering
\resizebox{0.4\hsize}{!}{\includegraphics{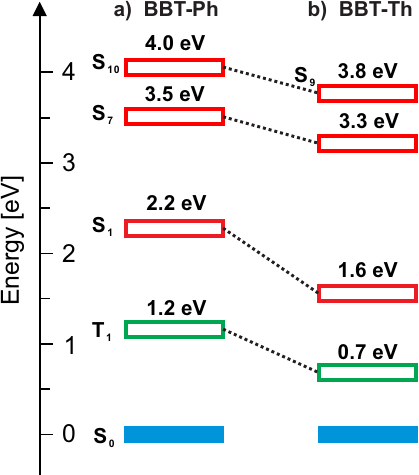}}
\caption{Electronic transitions determined with electronic HREELS at a) 11 ML BBT-Ph/Au(111) and  b) 5 ML BBT-Th/Au(111). Among the singlet states $S_{1}$, $S_{7}$, and $S_{10}$/$S_{9}$ the energies of the energetically lowest triplet state ($T_{1})$ have been determined.}
 \label{energy-diagram}
\end{figure}

Figure \ref{energy-diagram} a)
summarizes the electronic transitions determined with electronic HREELS of 11 ML BBT-Ph/Au(111). The optical gap observed in the thin film is in agreement with the data measured with UV/vis spectroscopy in solution (see Figure \ref{ele_BBT-Ph} b)), only a small energy-shift (50 meV) is found.  This suggests a relatively weak influence of the metal substrate (adsorbate/substrate interactions) and intermolecular forces (adsorbate/adsorbate interactions) on the electronic states in the BBT-Ph thin film.

\begin{figure}[H]
	\centering
    \begin{subfigure}{\textwidth}
    \centering
        \includegraphics[width=0.30\textwidth]{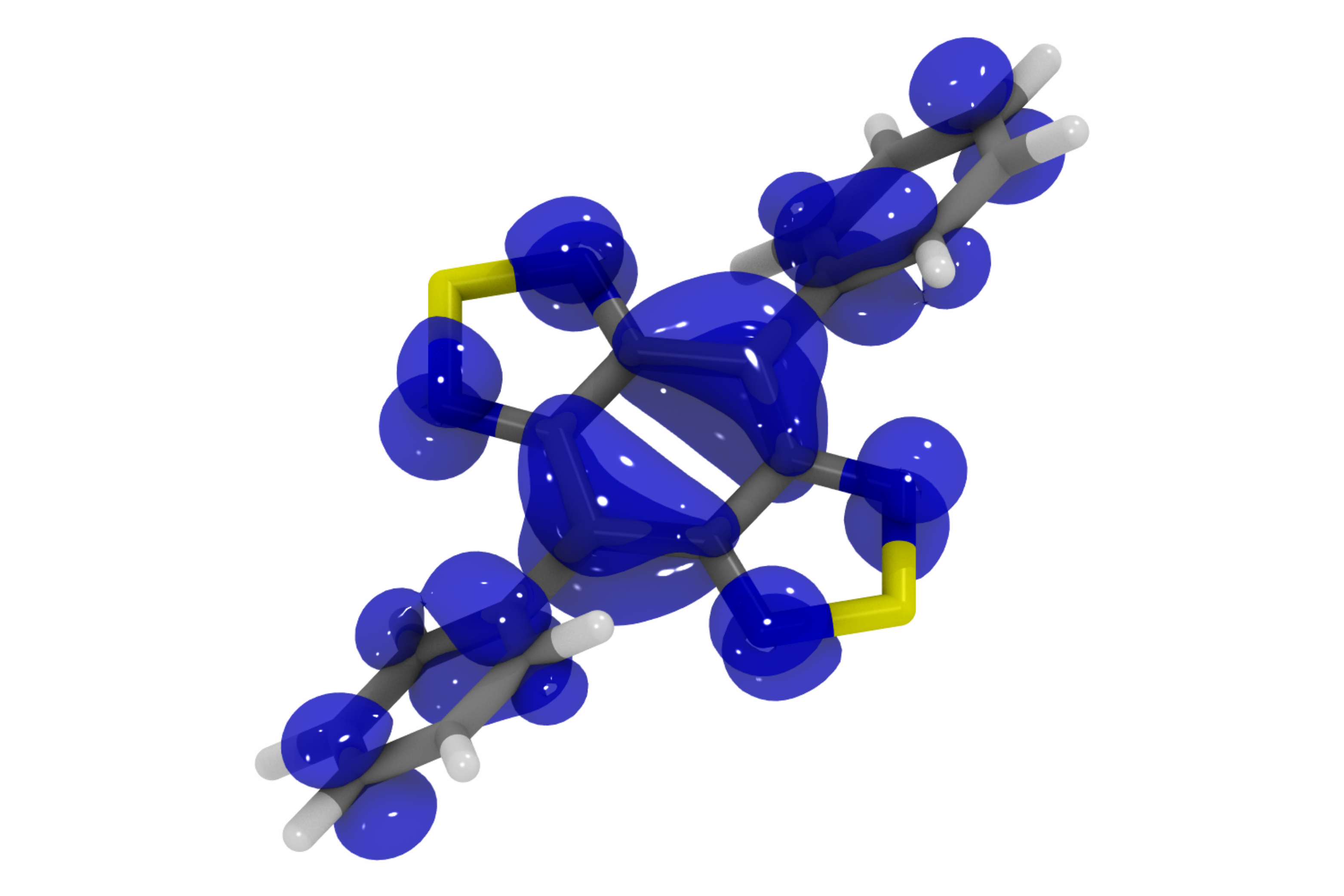}
        \includegraphics[width=0.30\textwidth]{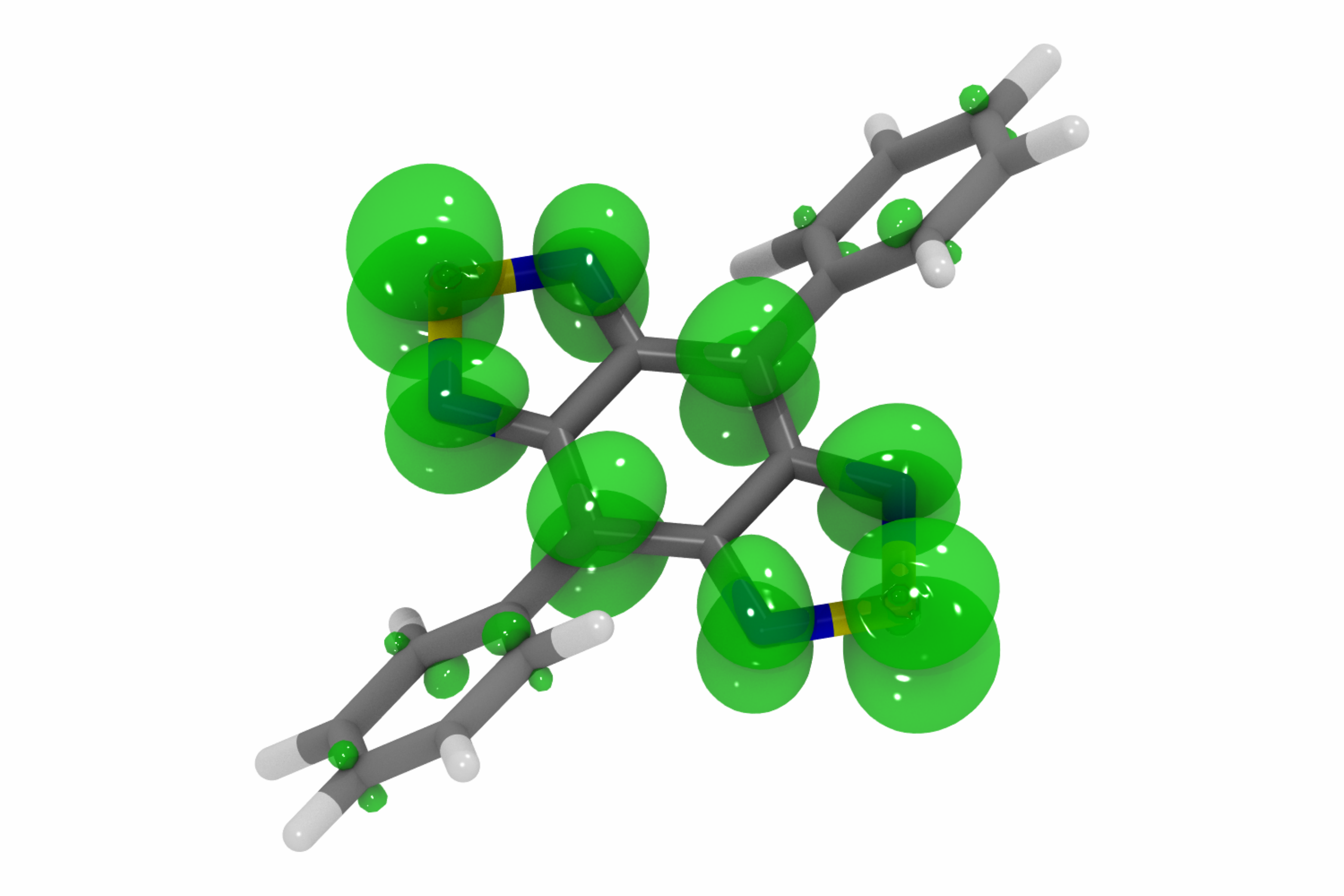}
        \subcaption{$\text{S}_0\rightarrow\text{S}_1$ detachment (left) and attachment density (right)}
    \end{subfigure}
    \begin{subfigure}{\textwidth}
    \centering
        \includegraphics[width=0.30\textwidth]{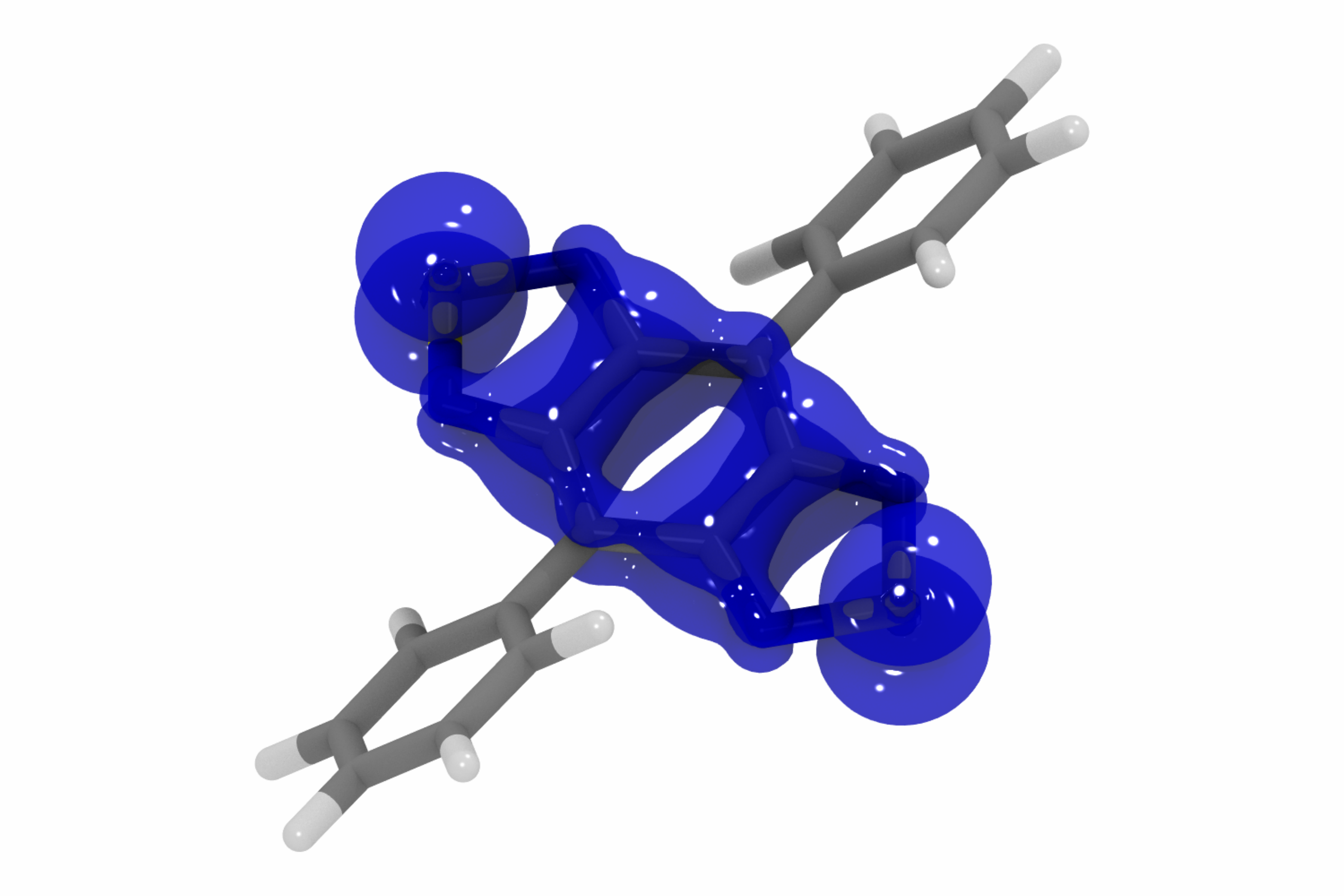}
        \includegraphics[width=0.30\textwidth]{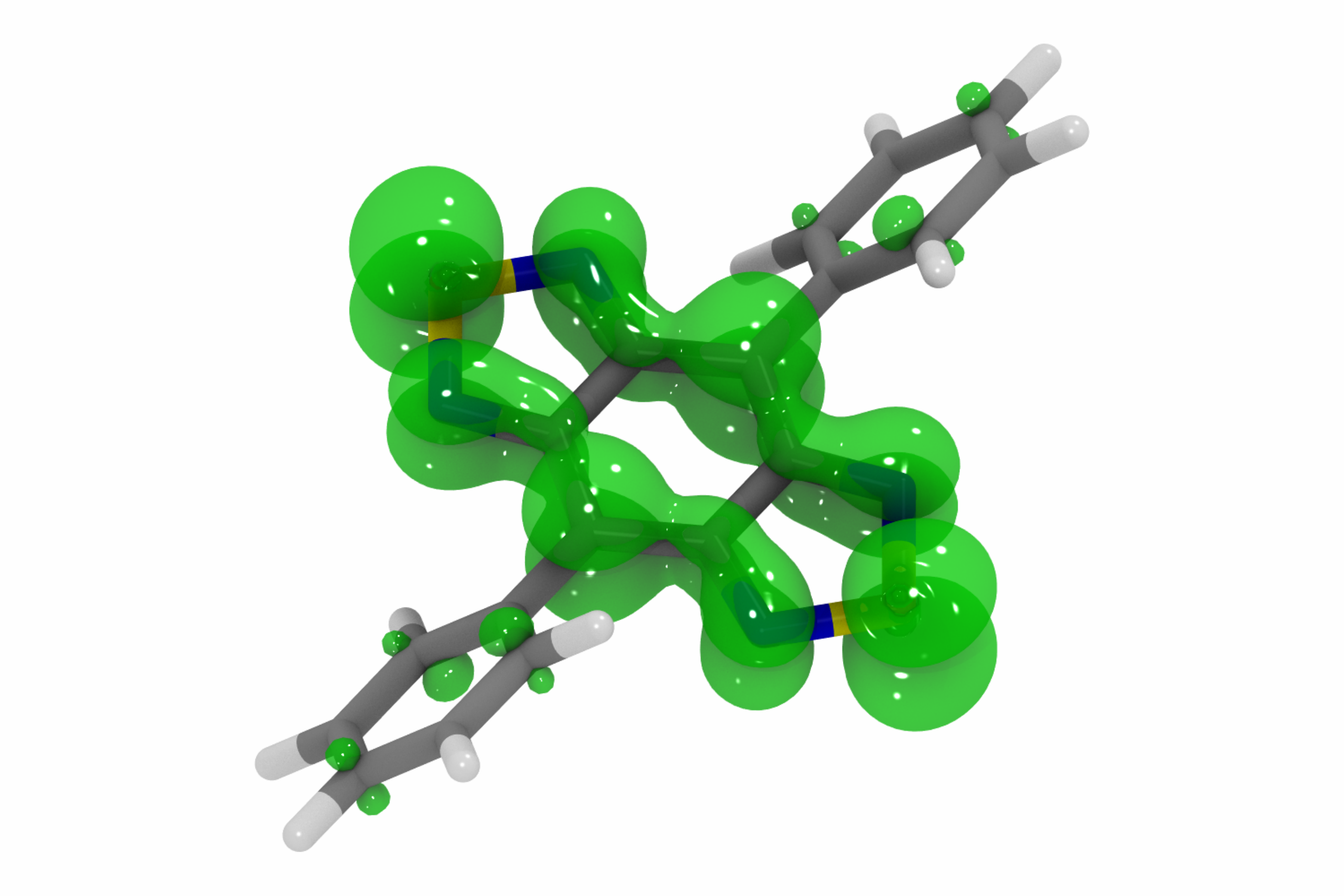}
        \subcaption{$\text{S}_0\rightarrow\text{S}_7$ detachment (left) and attachment density (right)}
    \end{subfigure}
    \begin{subfigure}{\textwidth}
    \centering
        \includegraphics[width=0.30\textwidth]{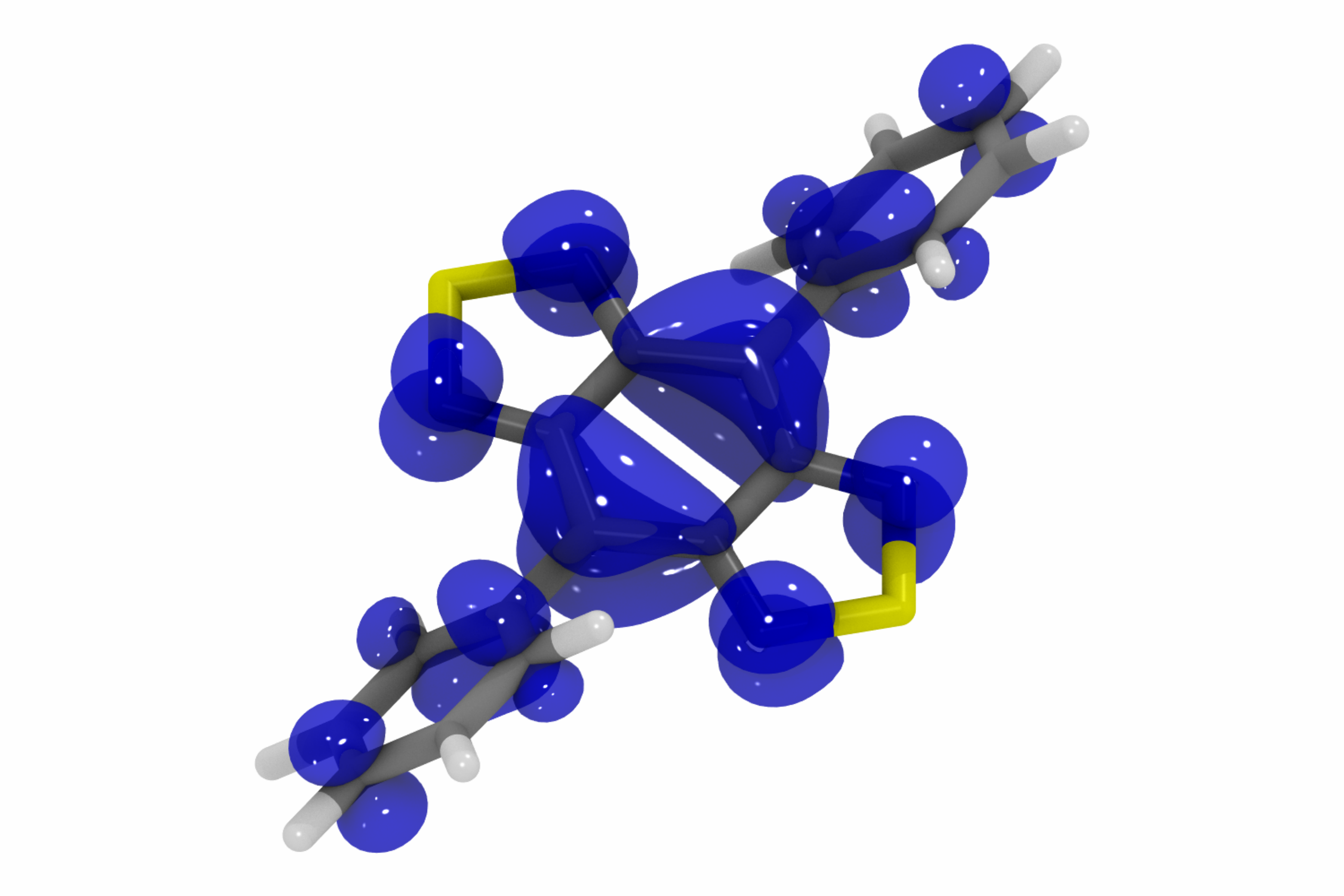}
        \includegraphics[width=0.30\textwidth]{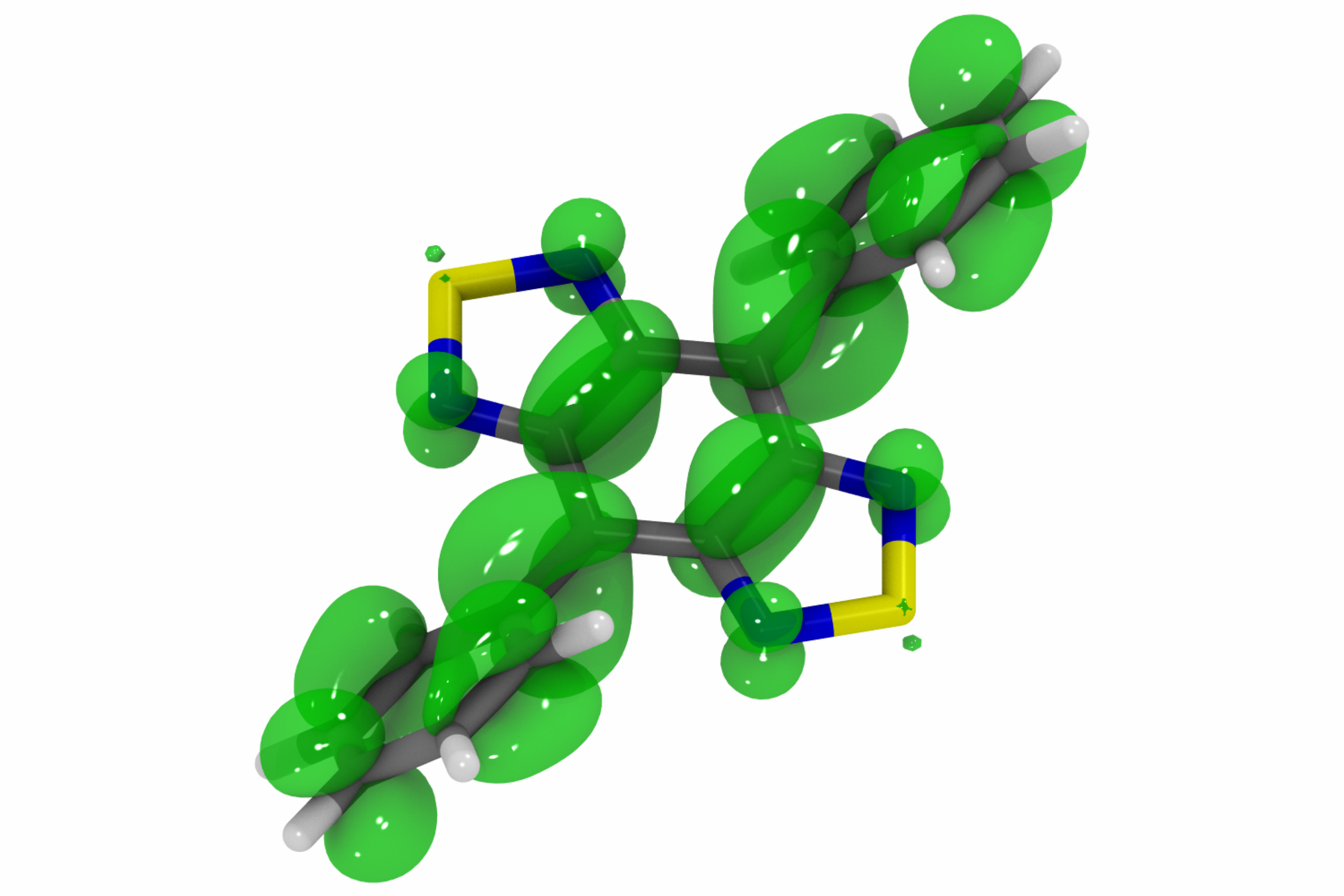}
        \subcaption{$\text{S}_0\rightarrow\text{S}_{10}$ detachment (left) and attachment density (right)}
    \end{subfigure}
	\caption{Detachment and attachment densities of the transitions of BBT-Ph given in \autoref{tab:electronical_transitions_bbt_ph}, obtained at the level of TDA/BMK/aug-cc-pVDZ, plotted with an isosurface value of 0.001.}
	\label{fig:bbt_attachment_detachment}
\end{figure}

On the basis of our quantum chemical calculations, the peak observed at 2.2 eV can be assigned to the $\text{S}_0\rightarrow\text{S}_{1}$ transition with
$\pi\pi^\ast$ character (see \autoref{fig:bbt_attachment_detachment} and \autoref{tab:electronical_transitions_bbt_ph}).
The transition around 3.5 eV corresponds to the $\text{S}_0\rightarrow\text{S}_{7}$ transition with $n\pi^\ast$ character involving the nitrogen lone pairs and the BBT $\pi$-system.
The excitation around 4.0 eV can be associated with the $\text{S}_0\rightarrow\text{S}_{10}$ transition with $\pi\pi^\ast$ character involving both the backbone and the phenyl substituent.
These assignments have been made for the equilibrium geometry of BBT with a torsion angle of 44.7$^\circ$ between the phenyl rings and the BBT backbone according to our calculations.
The UV/vis absorption spectrum has been obtained in solution, while the HREEL spectrum has been measured in a thin molecular film.
To account for these environmental effects the CPCM solvent model has been employed.
Furthermore, since the dielectric constant is likely higher in the multilayer compared to n-hexane solution, a higher dielectric constant, that of \ce{CHCl3} has been employed in our calculations.
However, the effect of these different dielectric constants within the CPCM model is very small (see supporting information, Fig. S8).
\begin{table}[h]
	\centering
	\caption{Assignment of the electronic transitions of BBT-Ph. Calculated transition energies (BMK/aug-cc-pVDZ) are compared to experimental values obtained with UV/vis (hexane) absorption spectroscopy and HREELS.}
	\label{tab:electronical_transitions_bbt_ph}
	\begin{tabular}{llccc}
		\hline
		\multirow{2}{*}{State} & \multirow{2}{*}{Character} & \multicolumn{3}{c}{Excitation energy [eV]} \\
				       & & calc. (\ce{CHCl3}) & exp. (UV/vis) & exp. (HREELS) \\
		\hline
		S$_1$ 		& $\pi\rightarrow\pi^\ast$ 	& 2.17 & 2.2 & 2.2 \\
		S$_7$ 		& $n\rightarrow\pi^\ast$ 	& 3.96 & 3.5 & 3.5 \\
		S$_{10}$ 	& $\pi\rightarrow\pi^\ast$ 	& 4.46 & 4.2 & 4.0 \\
		\hline
	\end{tabular}
\end{table}

\begin{figure}[htb]
\centering
\resizebox{0.45\hsize}{!}{\includegraphics{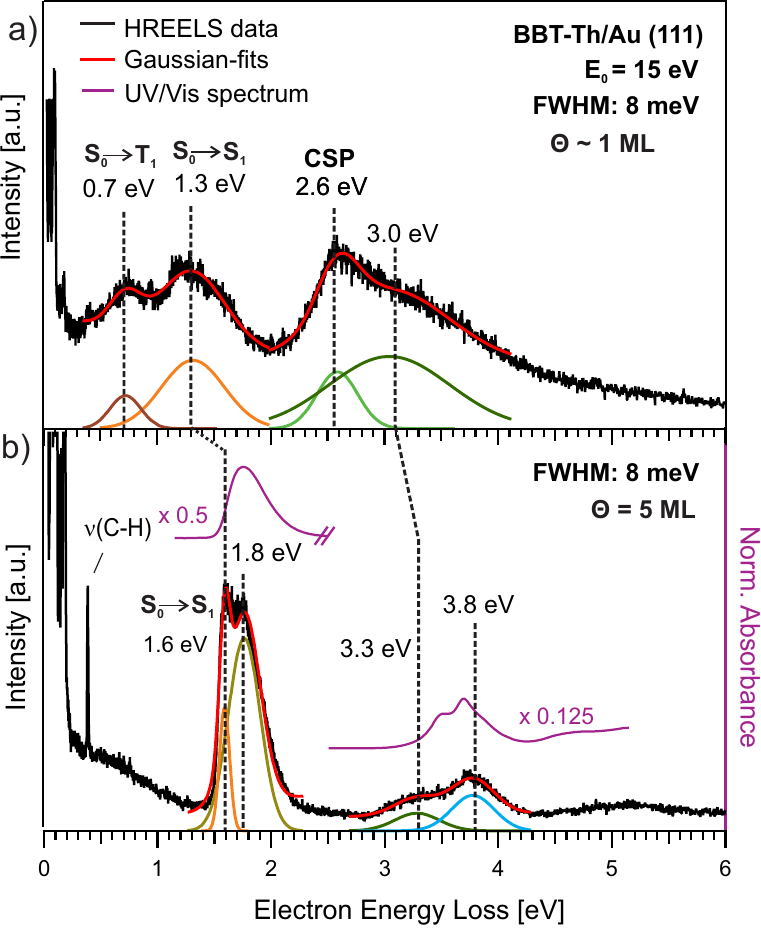}}
\caption{Electronic HREEL spectra of a)  1 ML and b) 5 ML BBT-Th/Au(111) measured with an incident electron energy of 15 eV and under specular-scattering geometry. The electronic transitions were fitted using Gaussian functions (red curves). The BBT-Th UV/vis spectrum obtained in hexane is shown for comparison (purple curve).}
\label{ele_BBT-Th}
\end{figure}

In a next step we studied the thiophene substituted BBT to elucidate the influence on the electronic structure compared to BBT-Ph.
Figure \ref{ele_BBT-Th} shows the monolayer and multilayer (5 ML) HREELS data.
In the monolayer regime four electronic transitions are detected located at 0.7, 1.3, 2.6, and 3.0 eV (peak maximum), which can be assigned to the first triplet state (0.7 eV),  the  $S_{0} \rightarrow S_{1}$ transition (optical gap, 1.3 eV), the CSP (2.6 eV) and $S_{0} \rightarrow S_{7}$  transition (3.0 eV) based on our calculations (see below).
 Compared to  BBT-Ph, an even more pronounced blue-shift of the transition energies are found when going from the mono- to the multilayer. The optical gap size rises by 300 meV from 1.3 to 1.6 eV and also by 300 eV for the $S_{0} \rightarrow S_{7}$ transition due to decoupling effects from the metal surface. Furthermore, in the multilayer the electron energy loss peak assigned to the optical gap shows apart from a pronounced  intensity increase, a vibronic contribution,  which is attributed to vibrational contributions of the symmetric  $\nu$(C--C) and $\nu$(C--N) stretching modes, the so-called breathing modes of the molecular backbone, with frequencies around 1450 $cm^{-1}$ (ca. 180 meV). In addition, a feature at 3.8 eV is detected, which can be assigned to $S_{0} \rightarrow S_{9}$ transition according to our calculations (see below). The comparison with the UV/vis data obtained from BBT-Th in solution shows an overall agreement. However, peak shifts around 200 meV towards lower energies are found in the thin film compared to the molecules in solution due to adsorbate/adsorbate (lateral) interactions. These interactions are obviously much stronger in BBT-Th in comparison to BBT-Ph, where we observed only a red-shift of around 50 meV.

\begin{figure}[H]
	\centering
    \begin{subfigure}{\textwidth}
    \centering
        \includegraphics[width=0.3\textwidth]{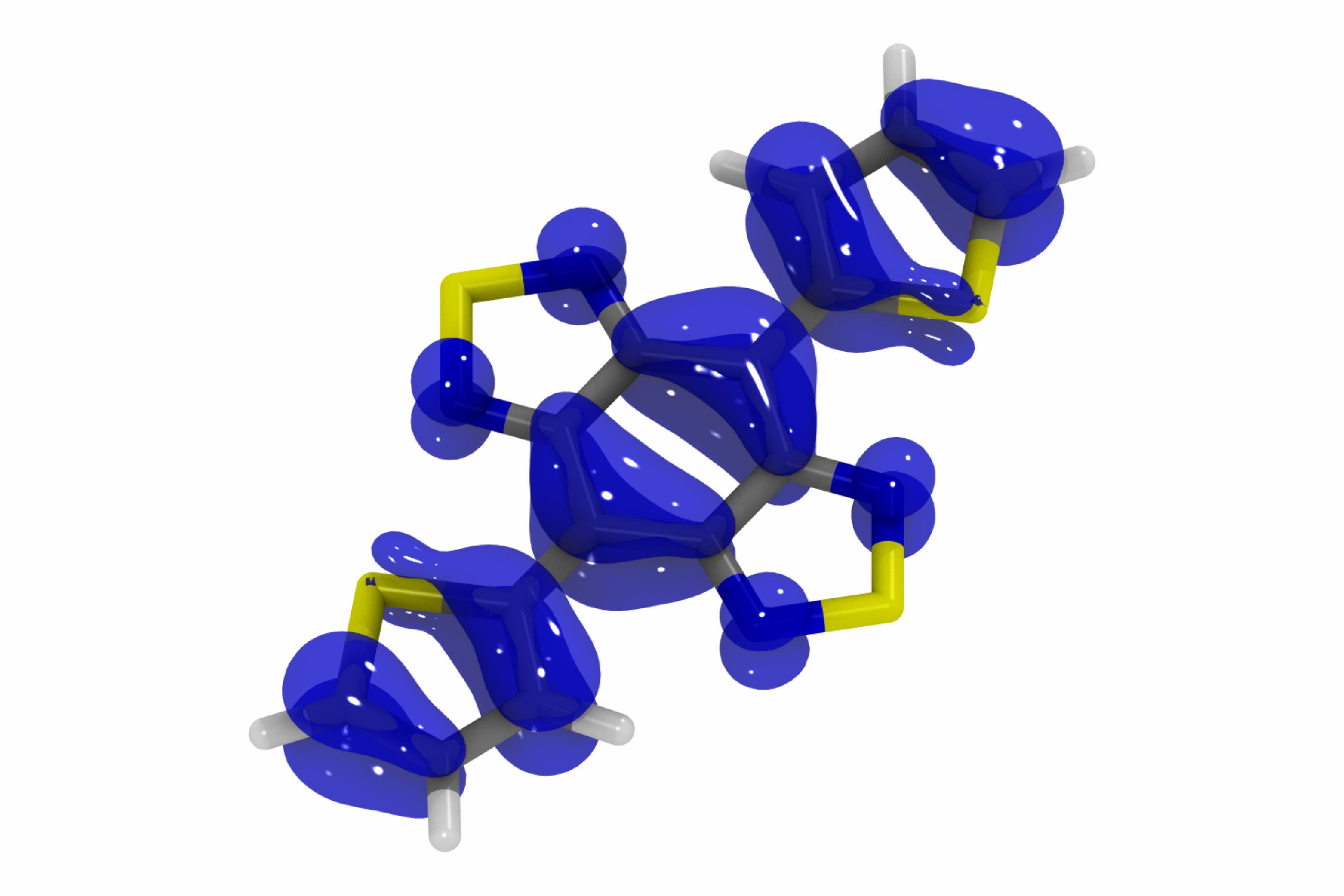}
        \includegraphics[width=0.3\textwidth]{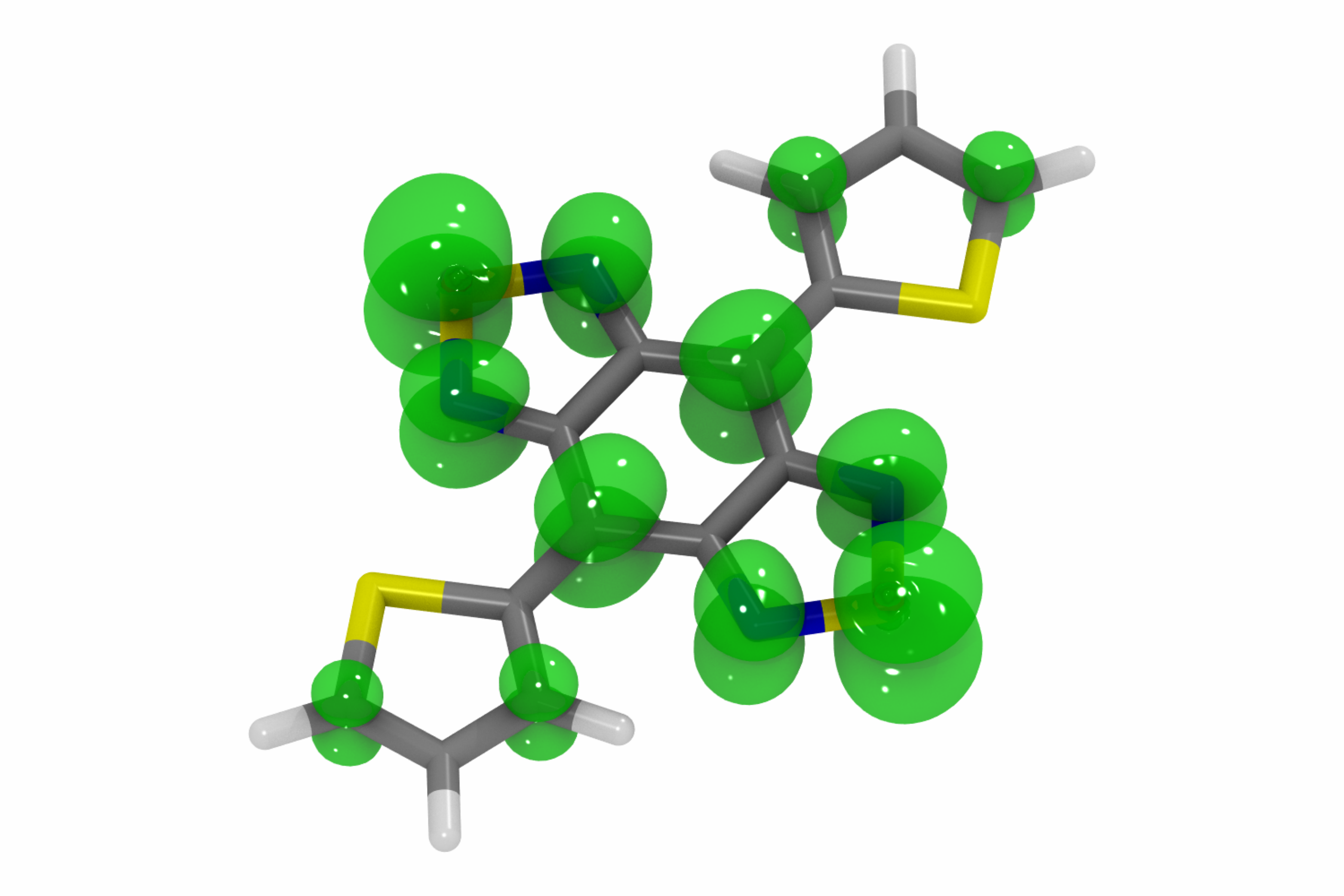}
        \subcaption{$\text{S}_0\rightarrow\text{S}_1$ detachment (left) and attachment density (right)}
    \end{subfigure}
    \begin{subfigure}{\textwidth}
    \centering
        \includegraphics[width=0.3\textwidth]{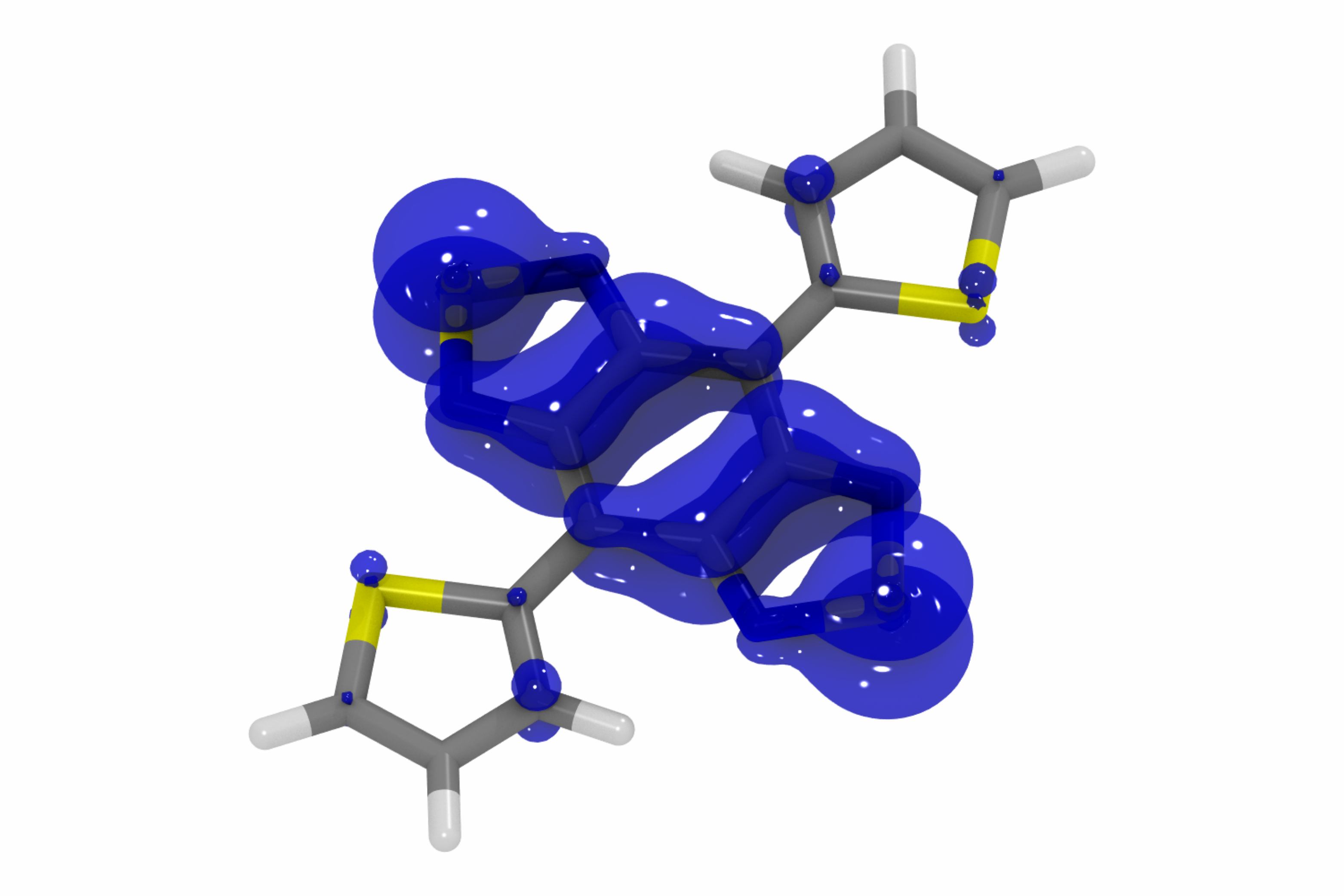}
        \includegraphics[width=0.3\textwidth]{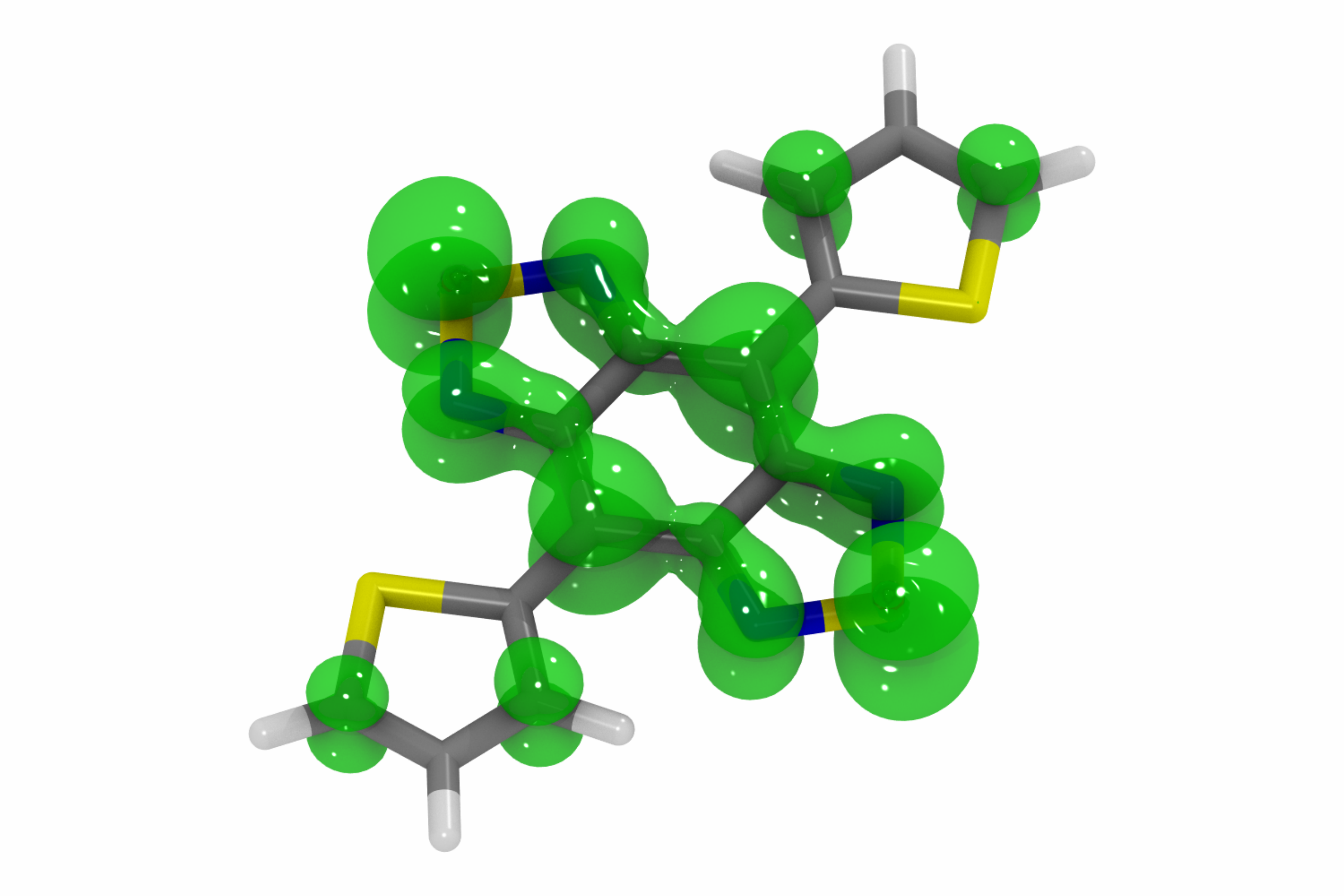}
        \subcaption{$\text{S}_0\rightarrow\text{S}_7$ detachment (left) and attachment density (right)}
    \end{subfigure}
    \begin{subfigure}{\textwidth}
    \centering
        \includegraphics[width=0.3\textwidth]{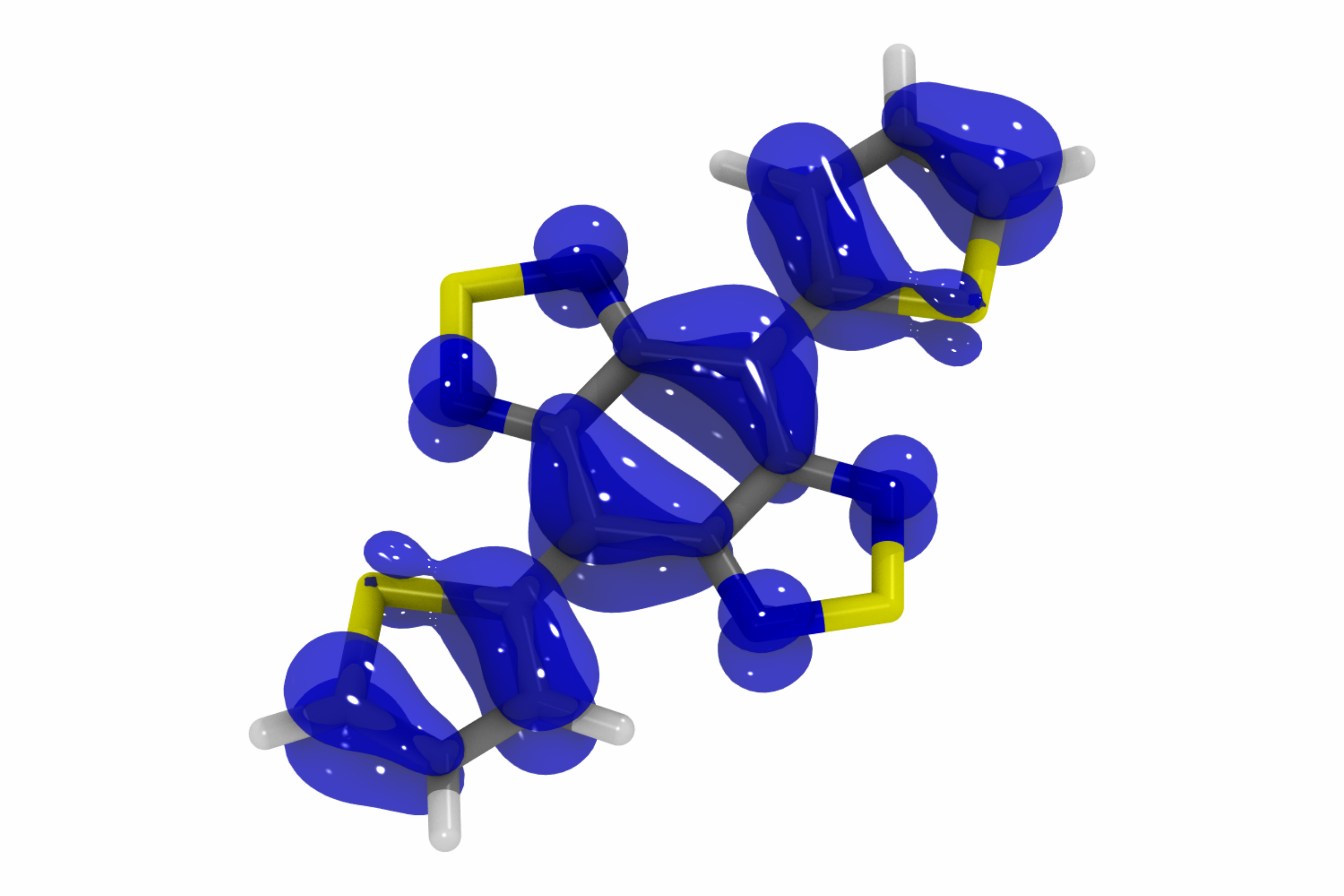}
        \includegraphics[width=0.3\textwidth]{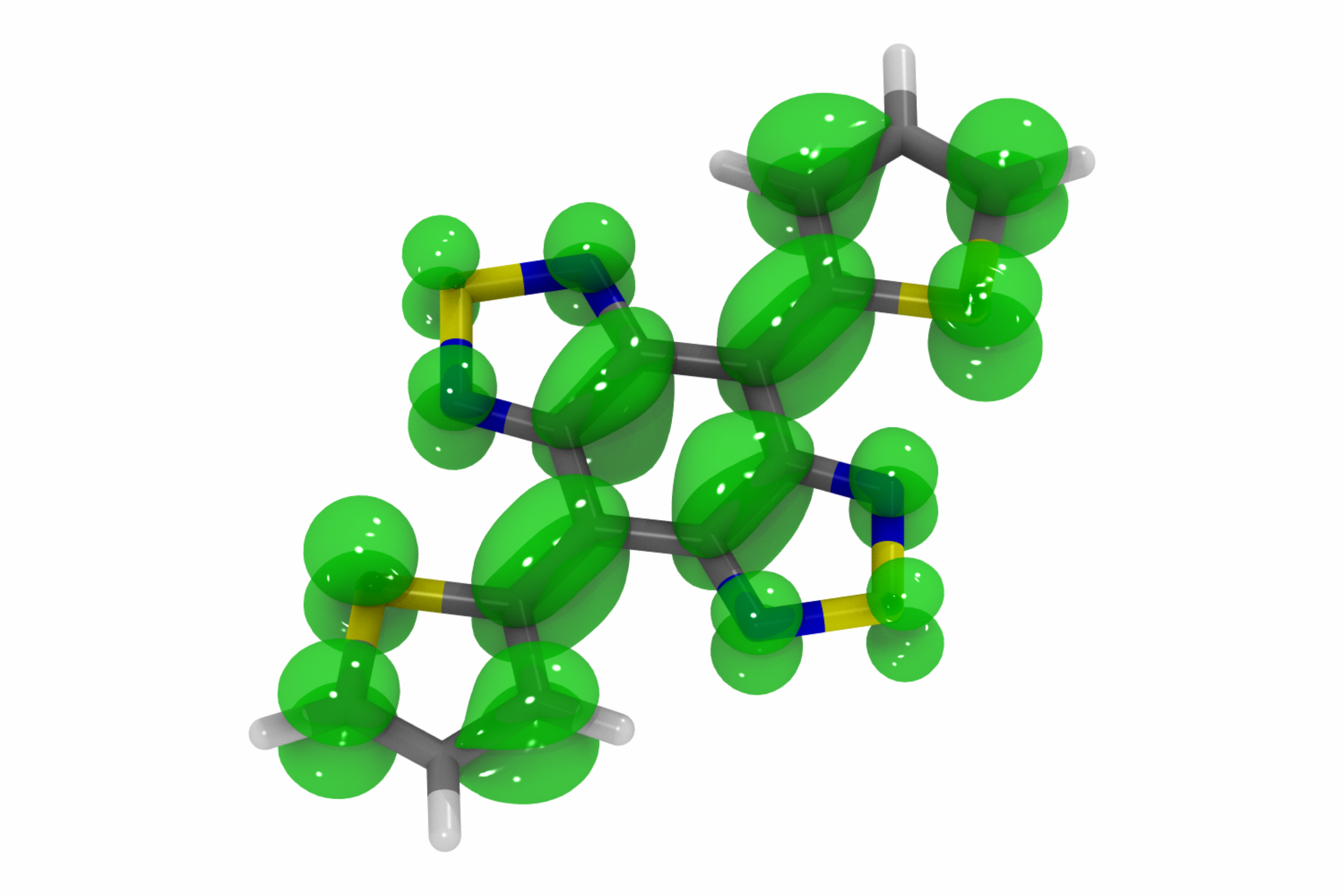}
        \subcaption{$\text{S}_0\rightarrow\text{S}_9$ detachment (left) and attachment density (right)}
    \end{subfigure}
	\caption{Detachment and attachment densities of the transitions of BBT-Th given in \autoref{tab:electronical_transitions_bbt_th}, obtained at the level of TDA/BMK/aug-cc-pVDZ, plotted with an isosurface value of 0.001.}
	\label{fig:bbt_th_attachment_detachment}
\end{figure}

According to our calculations, BBT-Th exhibits a planar equilibrium geometry in contrast to BBT-Ph.
The peak around 1.6 eV can be assigned to the $\text{S}_0\rightarrow\text{S}_1$ transition with $\pi\pi^\ast$ character involving both the BBT backbone and the thiophene substituent (see \autoref{fig:bbt_th_attachment_detachment} and  \autoref{tab:electronical_transitions_bbt_th}).
The observed peaks at 3.3 eV and 3.8 eV can be assigned to the $\text{S}_0\rightarrow\text{S}_7$ and $\text{S}_0\rightarrow\text{S}_9$ transitions with $\pi\pi^\ast$ character, respectively, of which the former is mostly centered at the BBT moiety while the latter involves both the backbone and substituent $\pi$-systems.
The $n\pi^\ast$ transition observed in BBT-Ph is a dark $\text{S}_{0}\rightarrow\text{S}_{10}$ excitation for BBT-Th (see supporting information).
This excitation is destabilized with respect to the $\pi\pi^\ast$ excitations in BBT-Th due to the interaction between the nitrogen and sulfur lone pairs within the molecular plane.
Additionally, it becomes a dark transition due to the orthogonality of the nitrogen $n$ orbitals and the $\pi$-system in the strictly planar geometry.

\begin{table}[h]
	\centering
	\caption{Assignment of the electronic transitions of BBT-Th. Calculated transition energies (TDA/BMK/aug-cc-pVDZ) are compared to experimental values obtained with UV/vis (hexane) absorption spectroscopy and HREELS.}
	\label{tab:electronical_transitions_bbt_th}

	\begin{tabular}{llccc}
		\hline
		\multirow{2}{*}{State} & \multirow{2}{*}{Character} & \multicolumn{3}{c}{Excitation energy [eV]} \\
				       & & calc. (\ce{CHCl3}) & exp. (UV/vis) & exp. (HREELS) \\
		\hline
		S$_1$ 		& $\pi\rightarrow\pi^\ast$ 	& 1.68 	& 1.9 & 1.6 \\
		S$_7$ 		& $\pi\rightarrow\pi^\ast$ 	& 3.91 	& 3.5 & 3.3 \\
		S$_{9}$ 	& $\pi\rightarrow\pi^\ast$ 	& 3.99 	& 3.7 & 3.8 \\
		\hline
	\end{tabular}
\end{table}

\begin{figure}[H]
	\centering
	\begin{subfigure}[b]{0.48\textwidth}
		\includegraphics[width=\linewidth]{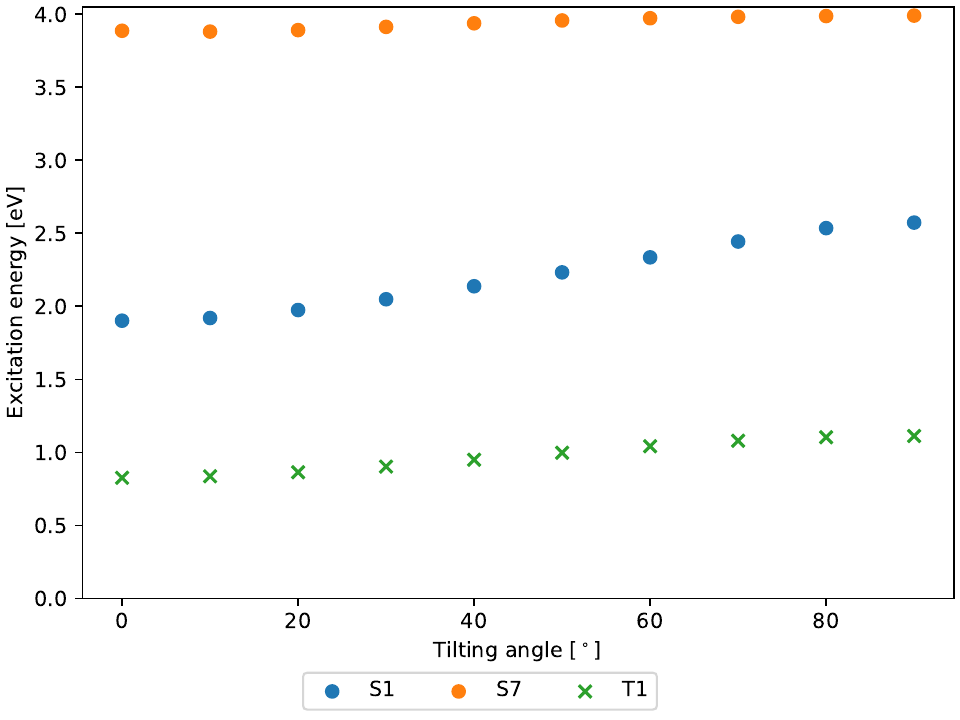}
		\caption{BBT-Ph}
		\label{fig:bbt_ph_angle_transition}
	\end{subfigure}
	\begin{subfigure}[b]{0.48\textwidth}
		\includegraphics[width=\linewidth]{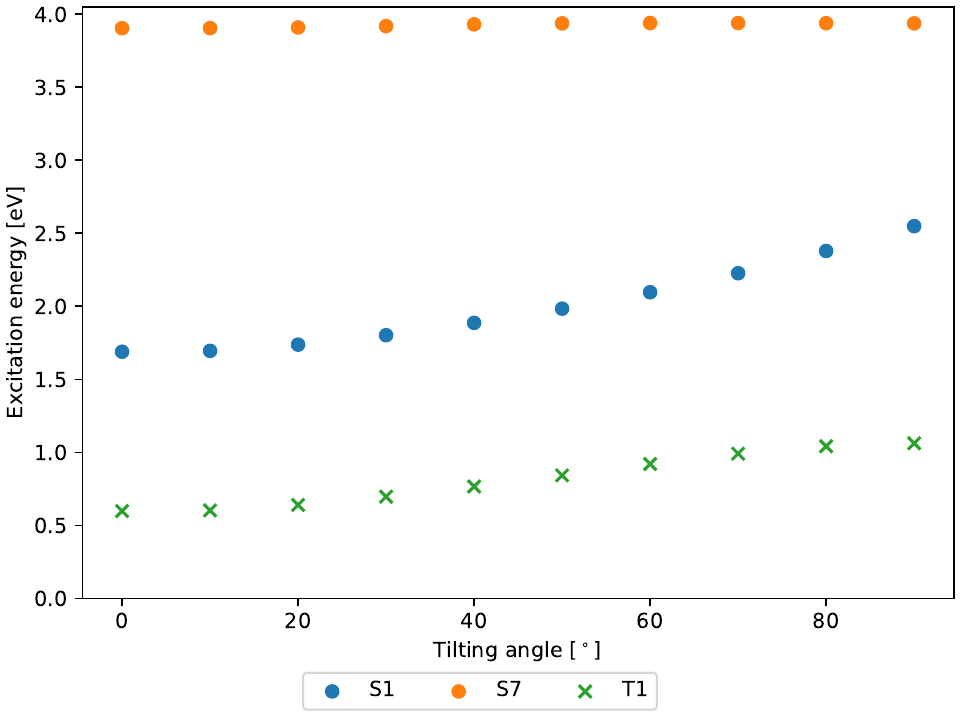}
		\caption{BBT-Th}
		\label{fig:bbt_th_angle_transition}
	\end{subfigure}
	\caption{Relationship between tilting angle and excited state energies. (\subref{fig:bbt_ph_angle_transition}) BBT-Ph (BMK/aug-cc-pVDZ, solvent: \ce{CHCl3}) optimized geometry has a tilting angle of $44.7^\circ$. (\subref{fig:bbt_th_angle_transition}) BBT-Th (BMK/aug-cc-pVDZ, solvent: DCM) optimized geometry is planar.
	The S$_7$ is taken from the relaxed geometry. For other angles the corresponding state with the same character as the S$_7$ is plotted.
}
	\label{fig:tilting_angle_transition_energy}
\end{figure}

Comparing the BBT-Ph and BBT-Th experimentally determined transition energies most strikingly all energies are reduced, the higher-lying singlet states by 200 meV ($S_{7}$ and $S_{10}$/$S_{9}$),  the optical gap ($\text{S}_0\rightarrow\text{S}_1$) even by 600 meV ($S_{1}$) and the first triplet state ($T_{1}$) energy by 500 meV (see Fig. \ref{energy-diagram}). Our calculated shift of the $\text{S}_0\rightarrow\text{S}_1$ transition energy between BBT-Ph and BBT-Th is 0.49 eV. According to our calculations, two factors contribute to this shift. The first one is the different angle between the substituent (Ph or Th) and the BBT moiety. While BBT-Ph exhibits a tilting angle of
44.7$^\circ$, BBT-Th is fully planar.
The relationship between the tilting angle and the excited state energies can be seen in \autoref{fig:tilting_angle_transition_energy}.
If both BBT-Ph and BBT-Th are in a planar geometry the difference in their $\text{S}_0\rightarrow\text{S}_1$ transition energy is reduced to only 0.25 eV.

If, however, an electron-withdrawing group, for example \ce{NO2} is computationally introduced at the thiophene ring, the corresponding $\text{S}_0\rightarrow\text{S}_1$ transition energy is increased to 1.95 eV which is very close to that of planar BBT-Ph.
This demonstrates the electron-rich nature of the five-membered thiophene ring is the second factor contributing to the red shift of the $\text{S}_0\rightarrow\text{S}_1$ transition in BBT-Th.
It should further be noted that the calculated $\text{S}_0\rightarrow\text{S}_1$ transition energy of 1.93 eV for planar BBT-Ph is in good agreement with the experimental value of 2.0 eV observed with HREELS when a single monolayer is measured, for which a planar geometry has been determined (see above).

For the T$_1$ state, an energy difference of 0.82 eV to the ground
state for planar BBT-Ph and one of 0.97 eV for an angle of 44.7$^\circ$ has been calculated.
Both of them are slightly lower than the experimentally observed one of 1.2 eV.
For BBT-Th, a computed T$_1$ energy of 0.6 eV is obtained which is in good agreement with the experimental value of 0.7 eV.

Again introducing an \ce{NO2} substituent at the thiophene increases the triplet state energy to 0.7 eV.
While still slightly lower than the T$_1$ energy of planar BBT-Ph (0.82 eV), it again indicates the importance of the electron-rich nature of the thiophene ring.
To further investigate the influence of the thiophene ring on the T$_1$ energy computationally, we replaced it by pyrrole.
Its relaxed geometry is again planar according to our calculations.
Furthermore, it exhibits nearly the same T$_1$ energy of 0.6 eV as BBT-Th, indicating that the sulfur atom is not the main reason for the low T$_1$ energy but rather the electron-rich nature of five-membered rings in conjunction with their preference for planar geometries.

The S$_7$ transition energy is calculated to be 3.96 eV for BBT-Ph and 3.91 eV for BBT-Th.
Both are blue shifted compared to the experimental value of around 3.5 eV (UV/vis in hexane).
Unlike the $\text{S}_0\rightarrow\text{S}_1$ transition energy, the higher lying transition is barely
affected by the tilting angle as seen in \autoref{fig:tilting_angle_transition_energy}, with the S$_7$ transition energy decreasing only by 0.08 eV when going from 0$^\circ$ to 90$^\circ$ tilting angle.

Compared to the HREELS value the S$_7$ transition energy is blue shifted by 0.2 eV in the
UV/vis experiment and 0.6 eV in the calculations. The difference in excitation energies between
BBT-Ph and BBT-Th in the HREELS experiment is not observed in either our calculations or the UV/vis spectra.
Neither of these takes the gold surface or intermolecular effects from the multilayer into account.
This indicates that the shift may originate from either the gold surface or interactions between the molecules within the layers.

\section{Conclusion}
Two benzobisthiadiazole (BBT) derivatives, a phenyl (BBT-Ph) and a thiophene (BBT-Th) core substituted compound adsorbed on Au(111) have been investigated using vibrational and electronic HREELS as well as quantum chemical calculations, focussing on the influence of core substitution pattern on the electronic structure. Both BBTs have been found to absorb in a planar adsorption geometry on Au(111), in which the molecular backbone as well as the Ph- or Th-substituent is oriented parallel to the Au surface in the mono- as well as multilayer regime. We were able to assign several features observed in the electronic HREEL spectrum to the corresponding excited electronic states based on TDDFT calculations. The optical gap ($S_{0} \rightarrow S_{1}$ transition) is found to be 2.2 eV for BBT-Ph and 1.6 eV for BBT-Th. Thus, the change in the substitution pattern from phenyl to thiophene has led to a pronounced reduction of the optical gap size by 0.6 eV. In addition, the energy of the first triplet state, which is located at 1.2 eV in BBT-Ph is also decreased by 0.5 eV in BBT-Th (T$_1$ = 0.7 eV). Our calculations clearly demonstrated that the reason for the reduced transition energies in BBT-Th is the electron-rich nature of the five-membered rings in conjunction with their preference for planar geometries. Using different substitution pattern may pave the way for fine-tuning the electronic properties of BBTs at will.

\section{Supporting information}
Temperature-programmed desorption data of BBT-Ph/Au(111) and BBT-Th/Au(111); additional HREELS data of BBT-Th/Au(111); scanning tunneling microscopy data of BBT-Th/Au(111); visualization of calculated vibrational modes of BBT-Ph and BBT-Th; calculated solvent effects on transition energies; calculated excitation energies and oscillator strengths

\section{Acknowledgments}

Funding by the German Research Foundation (DFG) through
the collaborative research center SFB 1249 "N-Heteropolycycles as Functional Materials" (project number 281029004-SFB 1249, projects A01, B01, and B06) is gratefully acknowledged.

\bibliography{BBT_lit}{}

\providecommand{\latin}[1]{#1}
\makeatletter
\providecommand{\doi}
  {\begingroup\let\do\@makeother\dospecials
  \catcode`\{=1 \catcode`\}=2 \doi@aux}
\providecommand{\doi@aux}[1]{\endgroup\texttt{#1}}
\makeatother
\providecommand*\mcitethebibliography{\thebibliography}
\csname @ifundefined\endcsname{endmcitethebibliography}
  {\let\endmcitethebibliography\endthebibliography}{}
\begin{mcitethebibliography}{89}
\providecommand*\natexlab[1]{#1}
\providecommand*\mciteSetBstSublistMode[1]{}
\providecommand*\mciteSetBstMaxWidthForm[2]{}
\providecommand*\mciteBstWouldAddEndPuncttrue
  {\def\EndOfBibitem{\unskip.}}
\providecommand*\mciteBstWouldAddEndPunctfalse
  {\let\EndOfBibitem\relax}
\providecommand*\mciteSetBstMidEndSepPunct[3]{}
\providecommand*\mciteSetBstSublistLabelBeginEnd[3]{}
\providecommand*\EndOfBibitem{}
\mciteSetBstSublistMode{f}
\mciteSetBstMaxWidthForm{subitem}{(\alph{mcitesubitemcount})}
\mciteSetBstSublistLabelBeginEnd
  {\mcitemaxwidthsubitemform\space}
  {\relax}
  {\relax}

\bibitem[Tam and Wu(2015)Tam, and Wu]{Tam2015}
Tam,~T. L.~D.; Wu,~J. Benzo[1,2-c:4,5-c']Bis[1,2,5]Thiadiazoles in Organic
  Optoelectronics: A Mini-Review. \emph{J. Mol. Eng. Mat.} \textbf{2015},
  \emph{3}, 15340003\relax
\mciteBstWouldAddEndPuncttrue
\mciteSetBstMidEndSepPunct{\mcitedefaultmidpunct}
{\mcitedefaultendpunct}{\mcitedefaultseppunct}\relax
\EndOfBibitem
\bibitem[Yuen \latin{et~al.}(2011)Yuen, Fan, Seifter, Lim, Hufschmid, Heeger,
  and Wudl]{Yuen2011}
Yuen,~J.~D.; Fan,~J.; Seifter,~J.; Lim,~B.; Hufschmid,~R.; Heeger,~A.~J.;
  Wudl,~F. High Performance Weak Donor–Acceptor Polymers in Thin Film
  Transistors: Effect of the Acceptor on Electronic Properties, Ambipolar
  Conductivity, Mobility, and Thermal Stability. \emph{J. Am. Chem. Soc.}
  \textbf{2011}, \emph{133}, 20799--20807\relax
\mciteBstWouldAddEndPuncttrue
\mciteSetBstMidEndSepPunct{\mcitedefaultmidpunct}
{\mcitedefaultendpunct}{\mcitedefaultseppunct}\relax
\EndOfBibitem
\bibitem[Ye \latin{et~al.}(2019)Ye, Chen, Pan, Liu, and Yin]{Ye2019}
Ye,~F.; Chen,~W.; Pan,~Y.; Liu,~S.~H.; Yin,~J. Benzobisthiadiazoles: From
  structure to function. \emph{Dyes and Pigments} \textbf{2019}, \emph{171},
  107746\relax
\mciteBstWouldAddEndPuncttrue
\mciteSetBstMidEndSepPunct{\mcitedefaultmidpunct}
{\mcitedefaultendpunct}{\mcitedefaultseppunct}\relax
\EndOfBibitem
\bibitem[Tam \latin{et~al.}(2012)Tam, Salim, Li, Zhou, Mhaisalkar, Su, Lam, and
  Grimsdale]{Tam2012}
Tam,~T. L.~D.; Salim,~T.; Li,~H.; Zhou,~F.; Mhaisalkar,~S.~G.; Su,~H.;
  Lam,~Y.~M.; Grimsdale,~A.~C. From benzobisthiadiazole, thiadiazoloquinoxaline
  to pyrazinoquinoxaline based polymers: effects of aromatic substituents on
  the performance of organic photovoltaics. \emph{J. Mater. Chem.}
  \textbf{2012}, \emph{22}, 18528--18534\relax
\mciteBstWouldAddEndPuncttrue
\mciteSetBstMidEndSepPunct{\mcitedefaultmidpunct}
{\mcitedefaultendpunct}{\mcitedefaultseppunct}\relax
\EndOfBibitem
\bibitem[M\"{u}ller \latin{et~al.}(2019)M\"{u}ller, Koser, Tverskoy, F,
  Freudenberg, and Bunz]{Muller2019}
M\"{u}ller,~M.; Koser,~S.; Tverskoy,~O.; F,~R.; Freudenberg,~J.; Bunz,~U. H.~F.
  Thiadiazolo-Azaacenes. \emph{Chem. Eur. J.} \textbf{2019}, \emph{25},
  6082--6086\relax
\mciteBstWouldAddEndPuncttrue
\mciteSetBstMidEndSepPunct{\mcitedefaultmidpunct}
{\mcitedefaultendpunct}{\mcitedefaultseppunct}\relax
\EndOfBibitem
\bibitem[Thomas and Bhanuprakash(2012)Thomas, and Bhanuprakash]{Thomas2012}
Thomas,~A.; Bhanuprakash,~K. Comparative Study of the Semiconducting Properties
  of Benzothiadiazole and Benzobis(thiadiazole) Derivatives Using Computational
  Techniques. \emph{Chem. Phys. Chem.} \textbf{2012}, \emph{13}, 597--605\relax
\mciteBstWouldAddEndPuncttrue
\mciteSetBstMidEndSepPunct{\mcitedefaultmidpunct}
{\mcitedefaultendpunct}{\mcitedefaultseppunct}\relax
\EndOfBibitem
\bibitem[Chmovzh \latin{et~al.}(2018)Chmovzh, Knyazeva, Mikhalchenko,
  Golovanov, Amelichev, and Rakitin]{Chmovzh2018}
Chmovzh,~T.~N.; Knyazeva,~E.~A.; Mikhalchenko,~L.~V.; Golovanov,~I.~S.;
  Amelichev,~S.~A.; Rakitin,~O.~A. Synthesis of the 4,7-Dibromo Derivative of
  Highly Electron-Deficient [1,2,5]Thiadiazolo[3,4-d]pyridazine and Its Cross-
  Coupling Reactions. \emph{Eur. J. Org. Chem.} \textbf{2018}, \emph{2018},
  5668--5677\relax
\mciteBstWouldAddEndPuncttrue
\mciteSetBstMidEndSepPunct{\mcitedefaultmidpunct}
{\mcitedefaultendpunct}{\mcitedefaultseppunct}\relax
\EndOfBibitem
\bibitem[Yamashita \latin{et~al.}(1997)Yamashita, Ono, Tomura, and
  Tanaka]{Yamashita1997}
Yamashita,~Y.; Ono,~K.; Tomura,~M.; Tanaka,~S. Synthesis and Properties of
  Benzobis(thiadiazole)s with Nonclassical $\pi$-Electron Ring Systems.
  \emph{Tetrahedron} \textbf{1997}, \emph{53}, 10178--10178\relax
\mciteBstWouldAddEndPuncttrue
\mciteSetBstMidEndSepPunct{\mcitedefaultmidpunct}
{\mcitedefaultendpunct}{\mcitedefaultseppunct}\relax
\EndOfBibitem
\bibitem[Qian and Wang(2010)Qian, and Wang]{Qian2010}
Qian,~G.; Wang,~Z.~Y. Near-Infrared Organic Compounds and Emerging
  Applications. \emph{Chem. As. J.} \textbf{2010}, \emph{5}, 1006--1029\relax
\mciteBstWouldAddEndPuncttrue
\mciteSetBstMidEndSepPunct{\mcitedefaultmidpunct}
{\mcitedefaultendpunct}{\mcitedefaultseppunct}\relax
\EndOfBibitem
\bibitem[Yuen \latin{et~al.}(2011)Yuen, Kumar, Zakhidov, Seifter, Lom, Heeger,
  and Wudl]{Yuen2011a}
Yuen,~J.~D.; Kumar,~R.; Zakhidov,~D.; Seifter,~J.; Lom,~B.; Heeger,~A.~J.;
  Wudl,~F. Ambipolarity in Benzobisthiadiazole-Based Donor–Acceptor
  Conjugated Polymers. \emph{Adv. Mater.} \textbf{2011}, \emph{23},
  3780--3785\relax
\mciteBstWouldAddEndPuncttrue
\mciteSetBstMidEndSepPunct{\mcitedefaultmidpunct}
{\mcitedefaultendpunct}{\mcitedefaultseppunct}\relax
\EndOfBibitem
\bibitem[Ishii \latin{et~al.}(1999)Ishii, Sugiyama, Ito, and Seki]{Ishii1999}
Ishii,~H.; Sugiyama,~K.; Ito,~E.; Seki,~K. Energy Level Alignment and
  Interfacial Electronic Structures at Organic/Metal and Organic/Organic
  Interfaces. \emph{Adv. Mater.} \textbf{1999}, \emph{11}, 605--625\relax
\mciteBstWouldAddEndPuncttrue
\mciteSetBstMidEndSepPunct{\mcitedefaultmidpunct}
{\mcitedefaultendpunct}{\mcitedefaultseppunct}\relax
\EndOfBibitem
\bibitem[Koch \latin{et~al.}(2013)Koch, Ueno, and Wee]{Koch2013}
Koch,~N., Ueno,~N., Wee,~A., Eds. \emph{The Molecule-Metal Interface};
  Wiley-VCH, Weinheim, 2013\relax
\mciteBstWouldAddEndPuncttrue
\mciteSetBstMidEndSepPunct{\mcitedefaultmidpunct}
{\mcitedefaultendpunct}{\mcitedefaultseppunct}\relax
\EndOfBibitem
\bibitem[Gruenewald \latin{et~al.}(2013)Gruenewald, Wachter, Meissner, Kozlik,
  Forker, and Fritz]{Gruenewald2013}
Gruenewald,~M.; Wachter,~K.; Meissner,~M.; Kozlik,~M.; Forker,~R.; Fritz,~T.
  Optical and Electronic Interaction at Metal--Organic and Organic--Organic
  Interfaces of Ultra-Thin Layers of PTCDA and SnPc on Noble Metal Surfaces.
  \emph{Org. Electron.} \textbf{2013}, \emph{14}, 2177--2183\relax
\mciteBstWouldAddEndPuncttrue
\mciteSetBstMidEndSepPunct{\mcitedefaultmidpunct}
{\mcitedefaultendpunct}{\mcitedefaultseppunct}\relax
\EndOfBibitem
\bibitem[Braun \latin{et~al.}(2009)Braun, Salaneck, and Fahlman]{Braun2009}
Braun,~S.; Salaneck,~W.~R.; Fahlman,~M. Energy-Level Alignment at Organic/Metal
  and Organic/Organic Interfaces. \emph{Adv. Mater.} \textbf{2009}, \emph{21},
  1450--1472\relax
\mciteBstWouldAddEndPuncttrue
\mciteSetBstMidEndSepPunct{\mcitedefaultmidpunct}
{\mcitedefaultendpunct}{\mcitedefaultseppunct}\relax
\EndOfBibitem
\bibitem[Oehzelt \latin{et~al.}(2014)Oehzelt, Koch, and Heimel]{Oehzelt2014}
Oehzelt,~M.; Koch,~N.; Heimel,~G. Organic semiconductor density of states
  controls the energy level alignment at electrode interfaces. \emph{Nature
  Comm.} \textbf{2014}, \emph{5}, 4174\relax
\mciteBstWouldAddEndPuncttrue
\mciteSetBstMidEndSepPunct{\mcitedefaultmidpunct}
{\mcitedefaultendpunct}{\mcitedefaultseppunct}\relax
\EndOfBibitem
\bibitem[Mercurio \latin{et~al.}(2013)Mercurio, Maurer, Liu, Hagen, Leyssner,
  Tegeder, Tkatchenko, Soubatch, Reuter, and Tautz]{Mercurio2013}
Mercurio,~G.; Maurer,~R.~J.; Liu,~W.; Hagen,~S.; Leyssner,~F.; Tegeder,~P.;
  Tkatchenko,~J. M.~A.; Soubatch,~S.; Reuter,~K.; Tautz,~F.~S. Quantification
  of Finite-Temperature Effects on Adsorption Geometries of $\pi$-Conjugated
  Molecules: Azobenzene/Ag(111). \emph{Phys. Rev. B} \textbf{2013}, \emph{88},
  035421\relax
\mciteBstWouldAddEndPuncttrue
\mciteSetBstMidEndSepPunct{\mcitedefaultmidpunct}
{\mcitedefaultendpunct}{\mcitedefaultseppunct}\relax
\EndOfBibitem
\bibitem[Bredas \latin{et~al.}(2009)Bredas, Norton, Cornil, and
  Coropceanu]{Bredas2009}
Bredas,~J.~L.; Norton,~J.~E.; Cornil,~J.; Coropceanu,~V. Molecular
  Understanding of Organic Solar Cells: The Challenges. \emph{Acc. Chem. Res.}
  \textbf{2009}, \emph{42}, 1691--1699\relax
\mciteBstWouldAddEndPuncttrue
\mciteSetBstMidEndSepPunct{\mcitedefaultmidpunct}
{\mcitedefaultendpunct}{\mcitedefaultseppunct}\relax
\EndOfBibitem
\bibitem[K\"{o}hler and B\"{a}ssler(2015)K\"{o}hler, and
  B\"{a}ssler]{Koehler2015}
K\"{o}hler,~A.; B\"{a}ssler,~H. \emph{Electronic Processes in Organic
  Semiconductors}; Wiley-VCH: Weinheim, Germany, 2015\relax
\mciteBstWouldAddEndPuncttrue
\mciteSetBstMidEndSepPunct{\mcitedefaultmidpunct}
{\mcitedefaultendpunct}{\mcitedefaultseppunct}\relax
\EndOfBibitem
\bibitem[May and K\"{u}hn(2011)May, and K\"{u}hn]{May2011}
May,~V.; K\"{u}hn,~O. \emph{Charge and Energy Transfer Dynamics in Molecular
  Systems}; Wiley-VCH: Weinheim, Germany, 2011\relax
\mciteBstWouldAddEndPuncttrue
\mciteSetBstMidEndSepPunct{\mcitedefaultmidpunct}
{\mcitedefaultendpunct}{\mcitedefaultseppunct}\relax
\EndOfBibitem
\bibitem[Beljonne \latin{et~al.}(2011)Beljonne, Cornil, Muccioli, Zannoni,
  Br\'{e}das, and Castet]{Beljonne2011}
Beljonne,~D.; Cornil,~J.; Muccioli,~L.; Zannoni,~C.; Br\'{e}das,~J.~L.;
  Castet,~F. Electronic Processes at Organic--Organic Interfaces: Insight from
  Modeling and Implications for Opto-electronic Devices. \emph{Chem. Mater.}
  \textbf{2011}, \emph{23}, 591--609\relax
\mciteBstWouldAddEndPuncttrue
\mciteSetBstMidEndSepPunct{\mcitedefaultmidpunct}
{\mcitedefaultendpunct}{\mcitedefaultseppunct}\relax
\EndOfBibitem
\bibitem[Ruhle \latin{et~al.}(2011)Ruhle, Lukyanov, May, Schrader, Vehoff,
  Kirkpatrick, Baumeier, and Andrienko]{Ruhle2011}
Ruhle,~V.; Lukyanov,~A.; May,~F.; Schrader,~M.; Vehoff,~T.; Kirkpatrick,~J.;
  Baumeier,~B.; Andrienko,~D. Microscopic Simulations of Charge Transport in
  Disordered Organic Semiconductors. \emph{J. Chem. Theory Comput.}
  \textbf{2011}, \emph{7}, 3335--3345\relax
\mciteBstWouldAddEndPuncttrue
\mciteSetBstMidEndSepPunct{\mcitedefaultmidpunct}
{\mcitedefaultendpunct}{\mcitedefaultseppunct}\relax
\EndOfBibitem
\bibitem[D$^\prime$Avino \latin{et~al.}(2016)D$^\prime$Avino, Muccioli, Castet,
  Poelking, Andrienko, Soos, Cornil, and Beljonne]{Avino2016}
D$^\prime$Avino,~G.; Muccioli,~L.; Castet,~F.; Poelking,~C.; Andrienko,~D.;
  Soos,~Z.~G.; Cornil,~J.; Beljonne,~D. Electrostatic Phenomena in Organic
  Semiconductors: Fundamentals and Implications for Photovoltaics. \emph{J.
  Phys.: Condens. Matter} \textbf{2016}, \emph{28}, 433002\relax
\mciteBstWouldAddEndPuncttrue
\mciteSetBstMidEndSepPunct{\mcitedefaultmidpunct}
{\mcitedefaultendpunct}{\mcitedefaultseppunct}\relax
\EndOfBibitem
\bibitem[Casu(2015)]{Casu2015}
Casu,~M.~B. Growth, Structure, and Electronic Properties in Organic Thin Films
  Deposited on Metal Surfaces Investigated by Low Energy Electron Microscopy
  and Photoelectron Emission Microscopy. \emph{J. Elect. Spec. Rel. Phenom.}
  \textbf{2015}, \emph{204}, 39--48\relax
\mciteBstWouldAddEndPuncttrue
\mciteSetBstMidEndSepPunct{\mcitedefaultmidpunct}
{\mcitedefaultendpunct}{\mcitedefaultseppunct}\relax
\EndOfBibitem
\bibitem[Forker \latin{et~al.}(2012)Forker, Gruenewald, and Fritz]{Forker2012}
Forker,~R.; Gruenewald,~M.; Fritz,~T. Optical Differential Reflectance
  Spectroscopy on Thin Molecular Films. \emph{Annu. Rep. Prog. Chem., Sect. C:
  Phys. Chem.} \textbf{2012}, \emph{108}, 34--68\relax
\mciteBstWouldAddEndPuncttrue
\mciteSetBstMidEndSepPunct{\mcitedefaultmidpunct}
{\mcitedefaultendpunct}{\mcitedefaultseppunct}\relax
\EndOfBibitem
\bibitem[Bronner \latin{et~al.}(2012)Bronner, Leyssner, S.Stremlau, Utecht,
  Saalfrank, Klamroth, and Tegeder]{Bronner2012}
Bronner,~C.; Leyssner,~F.; S.Stremlau,; Utecht,~M.; Saalfrank,~P.;
  Klamroth,~T.; Tegeder,~P. Electronic Structure of a Subnanometer wide
  Bottom-up Fabricated Graphene Nanoribbon: End States, Band Gap, and
  Dispersion. \emph{Phys. Rev. B: Condens. Matter Mater. Phys.} \textbf{2012},
  \emph{86}, 085444\relax
\mciteBstWouldAddEndPuncttrue
\mciteSetBstMidEndSepPunct{\mcitedefaultmidpunct}
{\mcitedefaultendpunct}{\mcitedefaultseppunct}\relax
\EndOfBibitem
\bibitem[Bronner \latin{et~al.}(2013)Bronner, Stremlau, Gille, Brau{\ss}e,
  Haase, Hecht, and Tegeder]{Bronner2013a}
Bronner,~C.; Stremlau,~S.; Gille,~M.; Brau{\ss}e,~F.; Haase,~A.; Hecht,~S.;
  Tegeder,~P. Aligning the Band Gap of Graphene Nanoribbons by Monomer Doping.
  \emph{Angew. Chem., Int. Ed.} \textbf{2013}, \emph{52}, 4422--4425\relax
\mciteBstWouldAddEndPuncttrue
\mciteSetBstMidEndSepPunct{\mcitedefaultmidpunct}
{\mcitedefaultendpunct}{\mcitedefaultseppunct}\relax
\EndOfBibitem
\bibitem[Maa{\ss} \latin{et~al.}(2017)Maa{\ss}, Utecht, Stremlau, Gille,
  Schwarz, Hecht, Klamroth, and Tegeder]{Maass2017}
Maa{\ss},~F.; Utecht,~M.; Stremlau,~S.; Gille,~M.; Schwarz,~J.; Hecht,~S.;
  Klamroth,~T.; Tegeder,~P. Electronic Structure changes during the on-surface
  synthesis of nitrogen-doped chevron-shaped graphene nanoribbons. \emph{Phys.
  Rev. B} \textbf{2017}, \emph{96}, 045434\relax
\mciteBstWouldAddEndPuncttrue
\mciteSetBstMidEndSepPunct{\mcitedefaultmidpunct}
{\mcitedefaultendpunct}{\mcitedefaultseppunct}\relax
\EndOfBibitem
\bibitem[Maass \latin{et~al.}(2016)Maass, Stein, Kohl, Hahn, Gade, Mastalerz,
  and Tegeder]{Maass2016}
Maass,~F.; Stein,~A.; Kohl,~B.; Hahn,~L.; Gade,~L.~H.; Mastalerz,~M.;
  Tegeder,~P. Substrate-Directed Growth of N-Heteropolycyclic Molecules on a
  Metal Surface. \emph{J. Phys. Chem. C} \textbf{2016}, \emph{120},
  2866--2873\relax
\mciteBstWouldAddEndPuncttrue
\mciteSetBstMidEndSepPunct{\mcitedefaultmidpunct}
{\mcitedefaultendpunct}{\mcitedefaultseppunct}\relax
\EndOfBibitem
\bibitem[Hahn \latin{et~al.}(2015)Hahn, Maass, Bleith, Zschieschang, Wadepohl,
  Klauk, Tegeder, and Gade]{maass2015}
Hahn,~L.; Maass,~F.; Bleith,~T.; Zschieschang,~U.; Wadepohl,~H.; Klauk,~H.;
  Tegeder,~P.; Gade,~L.~H. Core Halogenation as a Construction Principle in
  Tuning the Material Properties of Tetraazaperopyrenes. \emph{Chem. Eur. J.}
  \textbf{2015}, \emph{21}, 17691--17700\relax
\mciteBstWouldAddEndPuncttrue
\mciteSetBstMidEndSepPunct{\mcitedefaultmidpunct}
{\mcitedefaultendpunct}{\mcitedefaultseppunct}\relax
\EndOfBibitem
\bibitem[Stein \latin{et~al.}(2017)Stein, Maa{\ss}, and Tegeder]{Stein2017}
Stein,~A.; Maa{\ss},~F.; Tegeder,~P. Triisopropylsilylethynyl-Pentacene on
  {Au(111)}: Adsorption Properties, Electronic Structure, and Singlet Fission
  Dynamics. \emph{J. Phys. Chem. C} \textbf{2017}, \emph{121},
  18075--18083\relax
\mciteBstWouldAddEndPuncttrue
\mciteSetBstMidEndSepPunct{\mcitedefaultmidpunct}
{\mcitedefaultendpunct}{\mcitedefaultseppunct}\relax
\EndOfBibitem
\bibitem[Stremlau \latin{et~al.}(2017)Stremlau, Maass, and
  Tegeder]{Stremlau2017}
Stremlau,~S.; Maass,~F.; Tegeder,~P. Adsorption and switching properties of
  nitrospiropyran on Bi(114). \emph{J. Phys.: Condens Matter} \textbf{2017},
  \emph{29}, 314004\relax
\mciteBstWouldAddEndPuncttrue
\mciteSetBstMidEndSepPunct{\mcitedefaultmidpunct}
{\mcitedefaultendpunct}{\mcitedefaultseppunct}\relax
\EndOfBibitem
\bibitem[Gahl \latin{et~al.}(2013)Gahl, Brete, Leyssner, Koch, McNellis,
  Mielke, Carley, Grill, Reuter, Tegeder, and Weinelt]{Gahl2013}
Gahl,~C.; Brete,~D.; Leyssner,~F.; Koch,~M.; McNellis,~E.~R.; Mielke,~J.;
  Carley,~R.; Grill,~L.; Reuter,~K.; Tegeder,~P.; Weinelt,~M. Coverage- and
  Temperature-Controlled Isomerization of an Imine Derivative on Au(111).
  \emph{J. Am. Chem. Soc.} \textbf{2013}, \emph{135}, 4273--4281\relax
\mciteBstWouldAddEndPuncttrue
\mciteSetBstMidEndSepPunct{\mcitedefaultmidpunct}
{\mcitedefaultendpunct}{\mcitedefaultseppunct}\relax
\EndOfBibitem
\bibitem[Leyssner \latin{et~al.}(2010)Leyssner, Hagen, \'{O}v\'{a}ri, Dokic,
  Saalfrank, Peters, Hecht, Klamroth, and Tegeder]{Leyssner2010}
Leyssner,~F.; Hagen,~S.; \'{O}v\'{a}ri,~L.; Dokic,~J.; Saalfrank,~P.;
  Peters,~M.~V.; Hecht,~S.; Klamroth,~T.; Tegeder,~P. Photoisomerization
  ability of molecular switches adsorbed on Au(111): Comparison between
  azobenzene and stilbene derivatives. \emph{J. Phys. Chem. C} \textbf{2010},
  \emph{114}, 1231--1239\relax
\mciteBstWouldAddEndPuncttrue
\mciteSetBstMidEndSepPunct{\mcitedefaultmidpunct}
{\mcitedefaultendpunct}{\mcitedefaultseppunct}\relax
\EndOfBibitem
\bibitem[\'{O}v\'{a}ri \latin{et~al.}(2007)\'{O}v\'{a}ri, Wolf, and
  Tegeder]{Ovari2007}
\'{O}v\'{a}ri,~L.; Wolf,~M.; Tegeder,~P. Reversible Changes in the Vibrational
  Structure of of tetra-tert-butylazobenzene on a Au(111) Surface Induced by
  Light and Thermal Activation. \emph{J. Phys. Chem. C} \textbf{2007},
  \emph{111}, 15370--15374\relax
\mciteBstWouldAddEndPuncttrue
\mciteSetBstMidEndSepPunct{\mcitedefaultmidpunct}
{\mcitedefaultendpunct}{\mcitedefaultseppunct}\relax
\EndOfBibitem
\bibitem[Navarro \latin{et~al.}(2014)Navarro, Bocquet, Deperasinska, Pirug,
  Tautz, and Orrit]{Navarro2014}
Navarro,~P.; Bocquet,~F.~C.; Deperasinska,~I.; Pirug,~I.; Tautz,~F.~S.;
  Orrit,~M. Electron Energy Loss of Terrylene Deposited on Au(111): Vibrational
  and Electronic Spectroscopy. \emph{J. Phys. Chem. C} \textbf{2014},
  \emph{119}, 277--283\relax
\mciteBstWouldAddEndPuncttrue
\mciteSetBstMidEndSepPunct{\mcitedefaultmidpunct}
{\mcitedefaultendpunct}{\mcitedefaultseppunct}\relax
\EndOfBibitem
\bibitem[Ajdari \latin{et~al.}(2020)Ajdari, Stein, Hoffmann, M\"{u}ller, Bunz,
  Dreuw, and Tegeder]{Ajdari2020}
Ajdari,~M.; Stein,~A.; Hoffmann,~M.; M\"{u}ller,~M.; Bunz,~U. H.~F.; Dreuw,~A.;
  Tegeder,~P. Lightening up a Dark State of a Pentacene Derivative via
  N-Introduction. \emph{J. Phys. Chem. C} \textbf{2020}, \emph{124},
  7196--7204\relax
\mciteBstWouldAddEndPuncttrue
\mciteSetBstMidEndSepPunct{\mcitedefaultmidpunct}
{\mcitedefaultendpunct}{\mcitedefaultseppunct}\relax
\EndOfBibitem
\bibitem[Ajdari \latin{et~al.}(2021)Ajdari, Landwehr, Hoffmann, Hoffmann, Bunz,
  Dreuw, and Tegeder]{Ajdari2021}
Ajdari,~M.; Landwehr,~F.; Hoffmann,~M.; Hoffmann,~H.; Bunz,~U. H.~F.;
  Dreuw,~A.; Tegeder,~P. Influence of Core Halogenation on the Electronic
  Structure of Naphthothiadiazole Derivatives. \emph{J. Phys. Chem. C}
  \textbf{2021}, \emph{125}, 6359--6366\relax
\mciteBstWouldAddEndPuncttrue
\mciteSetBstMidEndSepPunct{\mcitedefaultmidpunct}
{\mcitedefaultendpunct}{\mcitedefaultseppunct}\relax
\EndOfBibitem
\bibitem[Ajdari \latin{et~al.}(2023)Ajdari, Pappenberger, Walla, Hoffmann,
  Michalsky, Kivala, Dreuw, and Tegeder]{Ajdari2023}
Ajdari,~M.; Pappenberger,~R.; Walla,~C.; Hoffmann,~M.; Michalsky,~I.;
  Kivala,~M.; Dreuw,~A.; Tegeder,~P. Impact of Connectivity on the Electronic
  Structure of N-Heterotriangulenes. \emph{J. Phys. Chem. C} \textbf{2023},
  \emph{127}, 542--549\relax
\mciteBstWouldAddEndPuncttrue
\mciteSetBstMidEndSepPunct{\mcitedefaultmidpunct}
{\mcitedefaultendpunct}{\mcitedefaultseppunct}\relax
\EndOfBibitem
\bibitem[Swiderek \latin{et~al.}(1990)Swiderek, Michaud, Hohlneicher, and
  Sanche]{Swiderek1990}
Swiderek,~P.; Michaud,~M.; Hohlneicher,~G.; Sanche,~L. Electron Energy Loss
  Spectroscopy of Solid Naphthalene and Acenaphthene: Search for the Low-Lying
  Triplet States. \emph{Chem. Phys. Lett.} \textbf{1990}, \emph{175},
  667--673\relax
\mciteBstWouldAddEndPuncttrue
\mciteSetBstMidEndSepPunct{\mcitedefaultmidpunct}
{\mcitedefaultendpunct}{\mcitedefaultseppunct}\relax
\EndOfBibitem
\bibitem[Swiderek \latin{et~al.}(1994)Swiderek, Fraser, Michaud, and
  Sanche]{Swiderek1994}
Swiderek,~P.; Fraser,~M.-J.; Michaud,~M.; Sanche,~L. Electron-Energy-Loss
  Spectroscopy of Low-Lying Triplet States of Styrene. \emph{J. Chem. Phys.}
  \textbf{1994}, \emph{100}, 70--77\relax
\mciteBstWouldAddEndPuncttrue
\mciteSetBstMidEndSepPunct{\mcitedefaultmidpunct}
{\mcitedefaultendpunct}{\mcitedefaultseppunct}\relax
\EndOfBibitem
\bibitem[Ajdari \latin{et~al.}(2020)Ajdari, Schmitt, Hoffmann, Maass, Reiss,
  Bunz, Dreuw, and Tegeder]{Ajdari2020b}
Ajdari,~M.; Schmitt,~T.; Hoffmann,~M.; Maass,~F.; Reiss,~H.; Bunz,~U. H.~F.;
  Dreuw,~A.; Tegeder,~P. Electronic Properties of 6, 13-Diazapentacene Adsorbed
  on Au(111): A Quantitative Determination of Transport, Singlet and Triplet
  States and Electronic Spectra. \emph{J. Phys. Chem. C} \textbf{2020},
  \emph{124}, 13196--13205\relax
\mciteBstWouldAddEndPuncttrue
\mciteSetBstMidEndSepPunct{\mcitedefaultmidpunct}
{\mcitedefaultendpunct}{\mcitedefaultseppunct}\relax
\EndOfBibitem
\bibitem[Hoffmann \latin{et~al.}(2022)Hoffmann, Ajdari, Landwehr, Tverskoy,
  Bunz, Dreuw, and Tegeder]{Hoffmann2022}
Hoffmann,~M.; Ajdari,~M.; Landwehr,~F.; Tverskoy,~O.; Bunz,~U. H.~F.;
  Dreuw,~A.; Tegeder,~P. Influence of N-introduction in pentacene on the
  electronic stracture and excited electronic states. \emph{Phys. Chem. Chem.
  Phys.} \textbf{2022}, \emph{24}, 3924--3932\relax
\mciteBstWouldAddEndPuncttrue
\mciteSetBstMidEndSepPunct{\mcitedefaultmidpunct}
{\mcitedefaultendpunct}{\mcitedefaultseppunct}\relax
\EndOfBibitem
\bibitem[Hill \latin{et~al.}(2000)Hill, Kahn, Soosb, and R.A.~Pascal]{Hill2000}
Hill,~I.; Kahn,~A.; Soosb,~Z.; R.A.~Pascal,~J. Charge-separation energy in
  films of $\pi$-conjugated organic molecules. \emph{Chem. Phys. Lett.}
  \textbf{2000}, \emph{327}, 181--188\relax
\mciteBstWouldAddEndPuncttrue
\mciteSetBstMidEndSepPunct{\mcitedefaultmidpunct}
{\mcitedefaultendpunct}{\mcitedefaultseppunct}\relax
\EndOfBibitem
\bibitem[Koch \latin{et~al.}(2005)Koch, Heimel, Wu, Zojer, Johnson, Br\'{e}das,
  M\"{u}llen, and Rabe]{Koch2005}
Koch,~N.; Heimel,~G.; Wu,~J.; Zojer,~E.; Johnson,~R.~L.; Br\'{e}das,~J.-L.;
  M\"{u}llen,~K.; Rabe,~J.~P. Influence of molecular conformation on
  organic/metal interface energetics. \emph{Chem Phys. Lett.} \textbf{2005},
  \emph{413}, 390--395\relax
\mciteBstWouldAddEndPuncttrue
\mciteSetBstMidEndSepPunct{\mcitedefaultmidpunct}
{\mcitedefaultendpunct}{\mcitedefaultseppunct}\relax
\EndOfBibitem
\bibitem[Kiguchi \latin{et~al.}(2004)Kiguchi, Entani, Saiki1, and
  Yoshikawa]{Kiguchi2004}
Kiguchi,~M.; Entani,~S.; Saiki1,~K.; Yoshikawa,~G. One-dimensional ordered
  structure of $\alpha$-sexithienyl on {C}u(110). \emph{Appl. Phys. Lett.}
  \textbf{2004}, \emph{84}, 3444--3446\relax
\mciteBstWouldAddEndPuncttrue
\mciteSetBstMidEndSepPunct{\mcitedefaultmidpunct}
{\mcitedefaultendpunct}{\mcitedefaultseppunct}\relax
\EndOfBibitem
\bibitem[Yokoyama \latin{et~al.}(2006)Yokoyama, Kurata, and
  Tanaka]{Yokoyama2006}
Yokoyama,~T.; Kurata,~S.; Tanaka,~S. Direct identification of conformational
  isomers of adsorbed oligothiophene on {C}u(100). \emph{J. Phys. Chem. B}
  \textbf{2006}, \emph{110}, 18130\relax
\mciteBstWouldAddEndPuncttrue
\mciteSetBstMidEndSepPunct{\mcitedefaultmidpunct}
{\mcitedefaultendpunct}{\mcitedefaultseppunct}\relax
\EndOfBibitem
\bibitem[Kakudate \latin{et~al.}(2006)Kakudate, Tsukamoto, Nakaya, and
  Nakayama]{Kakudate2006}
Kakudate,~T.; Tsukamoto,~S.; Nakaya,~M.; Nakayama,~T. Initial stage of
  adsorption of octithiophene molecules on {C}u(111). \emph{Surf. Sci.}
  \textbf{2006}, \emph{605}, 1021--1026\relax
\mciteBstWouldAddEndPuncttrue
\mciteSetBstMidEndSepPunct{\mcitedefaultmidpunct}
{\mcitedefaultendpunct}{\mcitedefaultseppunct}\relax
\EndOfBibitem
\bibitem[Kiel \latin{et~al.}(2007)Kiel, Duncker, Hagendorf, and
  Widdra]{Kiel2007}
Kiel,~M.; Duncker,~K.; Hagendorf,~C.; Widdra,~W. Molecular structure and chiral
  separation in $\alpha$-sexithiophene ultrathin films on {A}u(111): Low-energy
  electron diffraction and scanning tunneling microscopy. \emph{Phys. Rev. B}
  \textbf{2007}, \emph{75}, 195439\relax
\mciteBstWouldAddEndPuncttrue
\mciteSetBstMidEndSepPunct{\mcitedefaultmidpunct}
{\mcitedefaultendpunct}{\mcitedefaultseppunct}\relax
\EndOfBibitem
\bibitem[Grobosch and Knupfer(2007)Grobosch, and Knupfer]{Grobosch2007}
Grobosch,~M.; Knupfer,~M. Charge-Injection Barriers at Realistic Metal/Organic
  Interfaces: {M}etals Become Faceless. \emph{Adv. Mater.} \textbf{2007},
  \emph{19}, 754\relax
\mciteBstWouldAddEndPuncttrue
\mciteSetBstMidEndSepPunct{\mcitedefaultmidpunct}
{\mcitedefaultendpunct}{\mcitedefaultseppunct}\relax
\EndOfBibitem
\bibitem[Varene \latin{et~al.}(2011)Varene, Martin, and Tegeder]{Varene2011}
Varene,~E.; Martin,~I.; Tegeder,~P. Optically Induced Inter- and Intrafacial
  Electron Transfer Probed by Two-Photon Photoemission: Electronic States of
  Sexithiophene on Au(111). \emph{J. Phys. Chem. Lett.} \textbf{2011},
  \emph{2}, 252--256\relax
\mciteBstWouldAddEndPuncttrue
\mciteSetBstMidEndSepPunct{\mcitedefaultmidpunct}
{\mcitedefaultendpunct}{\mcitedefaultseppunct}\relax
\EndOfBibitem
\bibitem[Varene \latin{et~al.}(2012)Varene, Bogner, Bronner, and
  Tegeder]{Varene2012}
Varene,~E.; Bogner,~L.; Bronner,~C.; Tegeder,~P. Ultrafast exciton population,
  relaxation and decay dynamics in thin oligothiophene films. \emph{Phys. Rev.
  Lett.} \textbf{2012}, \emph{109}, 207601\relax
\mciteBstWouldAddEndPuncttrue
\mciteSetBstMidEndSepPunct{\mcitedefaultmidpunct}
{\mcitedefaultendpunct}{\mcitedefaultseppunct}\relax
\EndOfBibitem
\bibitem[Varene \latin{et~al.}(2012)Varene, Bogner, Meyer, Pennec, and
  Tegeder]{Varene2012a}
Varene,~E.; Bogner,~L.; Meyer,~S.; Pennec,~Y.; Tegeder,~P. Coverage-dependent
  adsorption geometry of octithiophene on Au(111). \emph{Phys. Chem. Chem.
  Phys.} \textbf{2012}, \emph{14}, 691--696\relax
\mciteBstWouldAddEndPuncttrue
\mciteSetBstMidEndSepPunct{\mcitedefaultmidpunct}
{\mcitedefaultendpunct}{\mcitedefaultseppunct}\relax
\EndOfBibitem
\bibitem[Bogner \latin{et~al.}(2015)Bogner, Yang, Corso, Fitzner, B{\"a}uerle,
  Franke, Pascual, and Tegeder]{bogner2015electronic}
Bogner,~L.; Yang,~Z.; Corso,~M.; Fitzner,~R.; B{\"a}uerle,~P.; Franke,~K.~J.;
  Pascual,~J.~I.; Tegeder,~P. Electronic Structure and Excited State Dynamics
  in a Dicyanovinyl-Substituted Oligothiophene on {Au(111)}. \emph{Phys. Chem.
  Chem- Phys.} \textbf{2015}, \emph{17}, 27118--27126\relax
\mciteBstWouldAddEndPuncttrue
\mciteSetBstMidEndSepPunct{\mcitedefaultmidpunct}
{\mcitedefaultendpunct}{\mcitedefaultseppunct}\relax
\EndOfBibitem
\bibitem[Bogner \latin{et~al.}(2016)Bogner, Yang, Baum, Corso, Fitzner,
  B{\"a}uerle, Franke, Pascual, and Tegeder]{bogner2016electronic}
Bogner,~L.; Yang,~Z.; Baum,~S.; Corso,~M.; Fitzner,~R.; B{\"a}uerle,~P.;
  Franke,~K.~J.; Pascual,~J.~I.; Tegeder,~P. Electronic States and Exciton
  Dynamics in Dicyanovinyl-Sexithiophene on {Au(111)}. \emph{J. Phys. Chem. C}
  \textbf{2016}, \emph{120}, 27268--27275\relax
\mciteBstWouldAddEndPuncttrue
\mciteSetBstMidEndSepPunct{\mcitedefaultmidpunct}
{\mcitedefaultendpunct}{\mcitedefaultseppunct}\relax
\EndOfBibitem
\bibitem[Bronsch \latin{et~al.}(2019)Bronsch, Wagner, Baum, Wansleben, Zielke,
  Ghanbari, Gy\"{o}r\"{o}k, Navarro-Quezada, Zeppenfeld, Weinelt, and
  Gahl]{Bronsch2019}
Bronsch,~W.; Wagner,~T.; Baum,~S.; Wansleben,~M.; Zielke,~K.; Ghanbari,~E.;
  Gy\"{o}r\"{o}k,~M.; Navarro-Quezada,~A.; Zeppenfeld,~P.; Weinelt,~M.;
  Gahl,~C. Interplay between Morphology and Electronic Structure in
  Sexithiophene Films on Au(111). \emph{J. Phys. Chem. C} \textbf{2019},
  \emph{123}, 7931--7939\relax
\mciteBstWouldAddEndPuncttrue
\mciteSetBstMidEndSepPunct{\mcitedefaultmidpunct}
{\mcitedefaultendpunct}{\mcitedefaultseppunct}\relax
\EndOfBibitem
\bibitem[Mishra \latin{et~al.}(2011)Mishra, Uhrich, Reinold, Pfeiffer, and
  B\"{a}uerle]{Mishra2011}
Mishra,~A.; Uhrich,~C.; Reinold,~E.; Pfeiffer,~M.; B\"{a}uerle,~P. Synthesis
  and Characterization of Acceptor-substituted Oligothiophenes for Solar Cell
  Applications. \emph{Adv. Energy Mater.} \textbf{2011}, \emph{1},
  265--273\relax
\mciteBstWouldAddEndPuncttrue
\mciteSetBstMidEndSepPunct{\mcitedefaultmidpunct}
{\mcitedefaultendpunct}{\mcitedefaultseppunct}\relax
\EndOfBibitem
\bibitem[Ziehlke \latin{et~al.}(2011)Ziehlke, Fitzner, K\"{o}rner, Gresser,
  Reinold, B\"{a}uerle, Leo, and Riede]{Ziehlke2011}
Ziehlke,~H.; Fitzner,~R.; K\"{o}rner,~C.; Gresser,~R.; Reinold,~E.;
  B\"{a}uerle,~P.; Leo,~K.; Riede,~M. Side Chain Variations on a Series of
  Dicyanovinyl-Terthiophenes: A Photoinduced Absorption Study. \emph{J. Phys.
  Chem. A} \textbf{2011}, \emph{115}, 8437--8446\relax
\mciteBstWouldAddEndPuncttrue
\mciteSetBstMidEndSepPunct{\mcitedefaultmidpunct}
{\mcitedefaultendpunct}{\mcitedefaultseppunct}\relax
\EndOfBibitem
\bibitem[Fitzner \latin{et~al.}(2011)Fitzner, Reinold, Mishra, Mena-Osteritz,
  Ziehlke, K\"{o}rner, Leo, Riede, Weil, Tsaryova, and et~al.]{Fitzner2011}
Fitzner,~R.; Reinold,~E.; Mishra,~A.; Mena-Osteritz,~E.; Ziehlke,~H.;
  K\"{o}rner,~C.; Leo,~K.; Riede,~M.; Weil,~M.; Tsaryova,~O.; et~al.,
  Dicyanovinyl-Substituted Oligothiophenes: Structure-Property Relationships
  and Application in Vacuum-Processed Small-Molecule Organic Solar Cells.
  \emph{Adv. Funct. Mater.} \textbf{2011}, \emph{21}, 897–910\relax
\mciteBstWouldAddEndPuncttrue
\mciteSetBstMidEndSepPunct{\mcitedefaultmidpunct}
{\mcitedefaultendpunct}{\mcitedefaultseppunct}\relax
\EndOfBibitem
\bibitem[Fitzner \latin{et~al.}(2012)Fitzner, Mena-Osteritz, Mishra, Schulz,
  Reinold, Weil, K\"{o}rner, Ziehlke, Elschner, Leo, and et~al.]{Fitzner2012}
Fitzner,~R.; Mena-Osteritz,~E.; Mishra,~A.; Schulz,~G.; Reinold,~E.; Weil,~M.;
  K\"{o}rner,~C.; Ziehlke,~H.; Elschner,~C.; Leo,~K.; et~al., Correlation of
  $\pi$-Conjugated Oligomer Structure with Film Morphology and Organic Solar
  Cell Performance. \emph{J. Am. Chem. Soc.} \textbf{2012}, \emph{134},
  11064--11067\relax
\mciteBstWouldAddEndPuncttrue
\mciteSetBstMidEndSepPunct{\mcitedefaultmidpunct}
{\mcitedefaultendpunct}{\mcitedefaultseppunct}\relax
\EndOfBibitem
\bibitem[Meerheim \latin{et~al.}(2014)Meerheim, K\"{o}rner, and
  Leo]{Meerheim2014}
Meerheim,~R.; K\"{o}rner,~C.; Leo,~K. Highly Efficient Organic Multi-Junction
  Solar Cells with a Thiophene Based Donor Material. \emph{Appl. Phys. Lett.}
  \textbf{2014}, \emph{105}, 063306\relax
\mciteBstWouldAddEndPuncttrue
\mciteSetBstMidEndSepPunct{\mcitedefaultmidpunct}
{\mcitedefaultendpunct}{\mcitedefaultseppunct}\relax
\EndOfBibitem
\bibitem[Stein \latin{et~al.}(2021)Stein, Rolf, Lotze, G\"{u}nther, Gade,
  Franke, and Tegeder]{Stein2021}
Stein,~A.; Rolf,~D.; Lotze,~C.; G\"{u}nther,~B.; Gade,~L.~H.; Franke,~K.~J.;
  Tegeder,~P. Band Formation at Interfaces Between N-Heteropolycycles and Gold
  Electrodes. \emph{J. Phys. Chem. Lett.} \textbf{2021}, \emph{12},
  947--951\relax
\mciteBstWouldAddEndPuncttrue
\mciteSetBstMidEndSepPunct{\mcitedefaultmidpunct}
{\mcitedefaultendpunct}{\mcitedefaultseppunct}\relax
\EndOfBibitem
\bibitem[Stein \latin{et~al.}(2021)Stein, Rolf, Lotze, Feldmann, Gerbert,
  Günther, Jeindl, Cartus, Hofmann, Gade, Franke, and Tegeder]{Stein2021b}
Stein,~A.; Rolf,~D.; Lotze,~C.; Feldmann,~S.; Gerbert,~D.; Günther,~B.;
  Jeindl,~A.; Cartus,~J.~J.; Hofmann,~O.~T.; Gade,~L.~H.; Franke,~K.~J.;
  Tegeder,~P. Electronic Properties of Tetraazaperopyrene Derivatives on
  Au(111): Energy Level Alignment and Interfacial Band Formation. \emph{J.
  Phys. Chem. C} \textbf{2021}, \emph{125}, 19969--19979\relax
\mciteBstWouldAddEndPuncttrue
\mciteSetBstMidEndSepPunct{\mcitedefaultmidpunct}
{\mcitedefaultendpunct}{\mcitedefaultseppunct}\relax
\EndOfBibitem
\bibitem[Klauk \latin{et~al.}(2007)Klauk, Zschieschang, Pflaum, and
  Halik]{klauk2007}
Klauk,~H.; Zschieschang,~U.; Pflaum,~J.; Halik,~M. Ultralow-power organic
  complementary circuits. \emph{Nature} \textbf{2007}, \emph{445},
  745--748\relax
\mciteBstWouldAddEndPuncttrue
\mciteSetBstMidEndSepPunct{\mcitedefaultmidpunct}
{\mcitedefaultendpunct}{\mcitedefaultseppunct}\relax
\EndOfBibitem
\bibitem[W{\"u}rthner and Stolte(2011)W{\"u}rthner, and Stolte]{wurthner2011}
W{\"u}rthner,~F.; Stolte,~M. Naphthalene and perylene diimides for organic
  transistors. \emph{Chem. Commun.} \textbf{2011}, \emph{47}, 5109--5115\relax
\mciteBstWouldAddEndPuncttrue
\mciteSetBstMidEndSepPunct{\mcitedefaultmidpunct}
{\mcitedefaultendpunct}{\mcitedefaultseppunct}\relax
\EndOfBibitem
\bibitem[Bunz \latin{et~al.}(2013)Bunz, Engelhart, Lindner, and
  Schaffroth]{Bunz2013}
Bunz,~U. H.~F.; Engelhart,~J.~U.; Lindner,~B.~D.; Schaffroth,~M. Large
  N-Heteroacenes: New Tricks for Very Old Dogs? \emph{Angew. Chem. Int. Ed.}
  \textbf{2013}, \emph{52}, 3810--3821\relax
\mciteBstWouldAddEndPuncttrue
\mciteSetBstMidEndSepPunct{\mcitedefaultmidpunct}
{\mcitedefaultendpunct}{\mcitedefaultseppunct}\relax
\EndOfBibitem
\bibitem[Miao(2014)]{miao2014ten}
Miao,~Q. {Ten Years of N-Heteropentacenes as Semiconductors for Organic
  Thin-Film Transistors}. \emph{Adv. Mater.} \textbf{2014}, \emph{26},
  5541--5549\relax
\mciteBstWouldAddEndPuncttrue
\mciteSetBstMidEndSepPunct{\mcitedefaultmidpunct}
{\mcitedefaultendpunct}{\mcitedefaultseppunct}\relax
\EndOfBibitem
\bibitem[Bunz(2015)]{bunz2015larger}
Bunz,~U. H.~F. {The larger linear N-heteroacenes}. \emph{Acc. Chem. Res.}
  \textbf{2015}, \emph{48}, 1676--1686\relax
\mciteBstWouldAddEndPuncttrue
\mciteSetBstMidEndSepPunct{\mcitedefaultmidpunct}
{\mcitedefaultendpunct}{\mcitedefaultseppunct}\relax
\EndOfBibitem
\bibitem[Li \latin{et~al.}(2011)Li, Tam, Lam, Mhaisalkar, and
  Grimsdale]{Li2011}
Li,~H.; Tam,~T.~L.; Lam,~Y.~M.; Mhaisalkar,~S.~G.; Grimsdale,~A.~C. Synthesis
  of Low Band Gap [1,2,5]-Thiadiazolo[3,4-g]quinoxaline Derivatives by
  Selective Reduction of Benzo[1,2-c;4,5-c’]bic[1,2,5]thiadiazole. \emph{Org.
  Lett.} \textbf{2011}, \emph{13}, 46--49\relax
\mciteBstWouldAddEndPuncttrue
\mciteSetBstMidEndSepPunct{\mcitedefaultmidpunct}
{\mcitedefaultendpunct}{\mcitedefaultseppunct}\relax
\EndOfBibitem
\bibitem[Ibach and Mills(1982)Ibach, and Mills]{ibach1982}
Ibach,~H.; Mills,~D.~L. \emph{Electron Energy Loss Spectroscopy and Surface
  Vibrations}; Academic Press, New York, 1982\relax
\mciteBstWouldAddEndPuncttrue
\mciteSetBstMidEndSepPunct{\mcitedefaultmidpunct}
{\mcitedefaultendpunct}{\mcitedefaultseppunct}\relax
\EndOfBibitem
\bibitem[Tegeder(2012)]{Tegeder2012}
Tegeder,~P. Optically and Thermally Induced Molecular Switching Processes at
  Metal Surfaces. \emph{J. Phys.: Condens. Matter} \textbf{2012}, \emph{24},
  394001\relax
\mciteBstWouldAddEndPuncttrue
\mciteSetBstMidEndSepPunct{\mcitedefaultmidpunct}
{\mcitedefaultendpunct}{\mcitedefaultseppunct}\relax
\EndOfBibitem
\bibitem[Frisch \latin{et~al.}(2009)Frisch, Trucks, Schlegel, Scuseria, Robb,
  Cheeseman, Scalmani, Barone, Mennucci, Petersson, Nakatsuji, Hada, M.~Ehara,
  Fukuda, Hasegawa, Ishida, Nakajima, Honda, Kitao, Nakai, Klene, Li, Knox,
  Hratchian, Cross, Bakken, Adamo, Jaramillo, Gomperts, Stratmann, Yazyev,
  Austin, Cammi, Pomelli, Ochterski, Ayala, Morokuma, Voth, Salvador,
  Dannenberg, Zakrzewski, Dapprich, Daniels, Strain, Farkas, Malick, Rabuck,
  Raghavachari, Foresman, Ortiz, Cui, Baboul, Clifford, J.~Cioslowski, Liu,
  Liashenko, Piskorz, Komaromi, Martin, Fox, Keith, Al-Laham, Peng,
  Nanayakkara, Challacombe, Gill, Johnson, Chen, Wong, Gonzalez, and
  Pople]{gaussian2009}
Frisch,~M. \latin{et~al.}  Gaussian 09, Revision A. 02, Gaussian. \emph{Inc.,
  Wallingford, CT} \textbf{2009}, \emph{200}\relax
\mciteBstWouldAddEndPuncttrue
\mciteSetBstMidEndSepPunct{\mcitedefaultmidpunct}
{\mcitedefaultendpunct}{\mcitedefaultseppunct}\relax
\EndOfBibitem
\bibitem[Hanwell \latin{et~al.}()Hanwell, Curtis, Lonie, Vandermeersch, Zurek,
  and Hutchison]{hanwell:2012:AvogadroAdvancedSemantic}
Hanwell,~M.~D.; Curtis,~D.~E.; Lonie,~D.~C.; Vandermeersch,~T.; Zurek,~E.;
  Hutchison,~G.~R. Avogadro: An Advanced Semantic Chemical Editor,
  Visualization, and Analysis Platform. \emph{4}, 17\relax
\mciteBstWouldAddEndPuncttrue
\mciteSetBstMidEndSepPunct{\mcitedefaultmidpunct}
{\mcitedefaultendpunct}{\mcitedefaultseppunct}\relax
\EndOfBibitem
\bibitem[Hohenberg and Kohn()Hohenberg, and
  Kohn]{hohenberg:1964:InhomogeneousElectronGasa}
Hohenberg,~P.; Kohn,~W. Inhomogeneous {{Electron Gas}}. \emph{136},
  B864--B871\relax
\mciteBstWouldAddEndPuncttrue
\mciteSetBstMidEndSepPunct{\mcitedefaultmidpunct}
{\mcitedefaultendpunct}{\mcitedefaultseppunct}\relax
\EndOfBibitem
\bibitem[Kohn and Sham()Kohn, and
  Sham]{kohn:1965:SelfConsistentEquationsIncluding}
Kohn,~W.; Sham,~L.~J. Self-{{Consistent Equations Including Exchange}} and
  {{Correlation Effects}}. \emph{140}, A1133--A1138\relax
\mciteBstWouldAddEndPuncttrue
\mciteSetBstMidEndSepPunct{\mcitedefaultmidpunct}
{\mcitedefaultendpunct}{\mcitedefaultseppunct}\relax
\EndOfBibitem
\bibitem[Becke()]{becke:1993:DensityFunctionalThermochemistry}
Becke,~A.~D. Density‐functional Thermochemistry. {{III}}. {{The}} Role of
  Exact Exchange. \emph{98}, 5648--5652\relax
\mciteBstWouldAddEndPuncttrue
\mciteSetBstMidEndSepPunct{\mcitedefaultmidpunct}
{\mcitedefaultendpunct}{\mcitedefaultseppunct}\relax
\EndOfBibitem
\bibitem[Woon and Dunning()Woon, and Dunning]{woon:1993:GaussianBasisSets}
Woon,~D.~E.; Dunning,~T.~H.,~Jr. Gaussian Basis Sets for Use in Correlated
  Molecular Calculations. {{III}}. {{The}} Atoms Aluminum through Argon.
  \emph{98}, 1358--1371\relax
\mciteBstWouldAddEndPuncttrue
\mciteSetBstMidEndSepPunct{\mcitedefaultmidpunct}
{\mcitedefaultendpunct}{\mcitedefaultseppunct}\relax
\EndOfBibitem
\bibitem[Kendall \latin{et~al.}()Kendall, Dunning, and
  Harrison]{kendall:1992:ElectronAffinitiesFirst}
Kendall,~R.~A.; Dunning,~T.~H.,~Jr.; Harrison,~R.~J. Electron Affinities of the
  First‐row Atoms Revisited. {{Systematic}} Basis Sets and Wave Functions.
  \emph{96}, 6796--6806\relax
\mciteBstWouldAddEndPuncttrue
\mciteSetBstMidEndSepPunct{\mcitedefaultmidpunct}
{\mcitedefaultendpunct}{\mcitedefaultseppunct}\relax
\EndOfBibitem
\bibitem[Dunning()]{dunning:1989:GaussianBasisSets}
Dunning,~T.~H.,~Jr. Gaussian Basis Sets for Use in Correlated Molecular
  Calculations. {{I}}. {{The}} Atoms Boron through Neon and Hydrogen.
  \emph{90}, 1007--1023\relax
\mciteBstWouldAddEndPuncttrue
\mciteSetBstMidEndSepPunct{\mcitedefaultmidpunct}
{\mcitedefaultendpunct}{\mcitedefaultseppunct}\relax
\EndOfBibitem
\bibitem[Caldeweyher \latin{et~al.}()Caldeweyher, Ehlert, Hansen, Neugebauer,
  Spicher, Bannwarth, and
  Grimme]{caldeweyherGenerallyApplicableAtomicCharge2019}
Caldeweyher,~E.; Ehlert,~S.; Hansen,~A.; Neugebauer,~H.; Spicher,~S.;
  Bannwarth,~C.; Grimme,~S. A {{Generally Applicable Atomic-Charge Dependent
  London Dispersion Correction Scheme}}.
  \url{https://chemrxiv.org/engage/chemrxiv/article-details/60c74060f96a006646286291}\relax
\mciteBstWouldAddEndPuncttrue
\mciteSetBstMidEndSepPunct{\mcitedefaultmidpunct}
{\mcitedefaultendpunct}{\mcitedefaultseppunct}\relax
\EndOfBibitem
\bibitem[Neese()]{neese:2018:SoftwareUpdateORCA}
Neese,~F. Software Update: The {{ORCA}} Program System, Version 4.0. \emph{8},
  e1327\relax
\mciteBstWouldAddEndPuncttrue
\mciteSetBstMidEndSepPunct{\mcitedefaultmidpunct}
{\mcitedefaultendpunct}{\mcitedefaultseppunct}\relax
\EndOfBibitem
\bibitem[Barone and Cossi()Barone, and
  Cossi]{barone:1998:QuantumCalculationMolecular}
Barone,~V.; Cossi,~M. Quantum {{Calculation}} of {{Molecular Energies}} and
  {{Energy Gradients}} in {{Solution}} by a {{Conductor Solvent Model}}.
  \emph{102}, 1995--2001\relax
\mciteBstWouldAddEndPuncttrue
\mciteSetBstMidEndSepPunct{\mcitedefaultmidpunct}
{\mcitedefaultendpunct}{\mcitedefaultseppunct}\relax
\EndOfBibitem
\bibitem[Hirata and Head-Gordon()Hirata, and
  Head-Gordon]{hirata:1999:TimedependentDensityFunctional}
Hirata,~S.; Head-Gordon,~M. Time-Dependent Density Functional Theory within the
  {{Tamm}}–{{Dancoff}} Approximation. \emph{314}, 291--299\relax
\mciteBstWouldAddEndPuncttrue
\mciteSetBstMidEndSepPunct{\mcitedefaultmidpunct}
{\mcitedefaultendpunct}{\mcitedefaultseppunct}\relax
\EndOfBibitem
\bibitem[Dreuw and Head-Gordon()Dreuw, and
  Head-Gordon]{dreuw:2005:SingleReferenceInitioMethods}
Dreuw,~A.; Head-Gordon,~M. Single-{{Reference}} Ab {{Initio Methods}} for the
  {{Calculation}} of {{Excited States}} of {{Large Molecules}}. \emph{105},
  4009--4037\relax
\mciteBstWouldAddEndPuncttrue
\mciteSetBstMidEndSepPunct{\mcitedefaultmidpunct}
{\mcitedefaultendpunct}{\mcitedefaultseppunct}\relax
\EndOfBibitem
\bibitem[Herbert()]{herbert:2023:DensityFunctionalTheory}
Herbert,~J.~M. \emph{Density {{Functional Theory}} for {{Electronic Excited
  States}}}; pp 69--118\relax
\mciteBstWouldAddEndPuncttrue
\mciteSetBstMidEndSepPunct{\mcitedefaultmidpunct}
{\mcitedefaultendpunct}{\mcitedefaultseppunct}\relax
\EndOfBibitem
\bibitem[Boese and Martin()Boese, and
  Martin]{boese:2004:DevelopmentDensityFunctionals}
Boese,~A.~D.; Martin,~J. M.~L. Development of Density Functionals for
  Thermochemical Kinetics. \emph{121}, 3405--3416\relax
\mciteBstWouldAddEndPuncttrue
\mciteSetBstMidEndSepPunct{\mcitedefaultmidpunct}
{\mcitedefaultendpunct}{\mcitedefaultseppunct}\relax
\EndOfBibitem
\bibitem[Kato \latin{et~al.}(2002)Kato, Noh, Hara, and Kawai]{Kato2002}
Kato,~H.~S.; Noh,~J.; Hara,~M.; Kawai,~M. An HREELS Study of Alkanethiol
  Self-Assembled Monolayers on Au(111). \emph{J. Phys. Chem. B} \textbf{2002},
  \emph{106}, 9655--9658\relax
\mciteBstWouldAddEndPuncttrue
\mciteSetBstMidEndSepPunct{\mcitedefaultmidpunct}
{\mcitedefaultendpunct}{\mcitedefaultseppunct}\relax
\EndOfBibitem
\bibitem[Park and Palmer(2009)Park, and Palmer]{Park2009}
Park,~S.~J.; Palmer,~R.~E. Plasmon Dispersion of the Au(111) Surface with and
  without Self-Assembled Monolayers. \emph{Phys. Rev. Lett.} \textbf{2009},
  \emph{102}, 216805\relax
\mciteBstWouldAddEndPuncttrue
\mciteSetBstMidEndSepPunct{\mcitedefaultmidpunct}
{\mcitedefaultendpunct}{\mcitedefaultseppunct}\relax
\EndOfBibitem
\bibitem[Park and Palmer(2010)Park, and Palmer]{Park2010}
Park,~S.~J.; Palmer,~R.~E. Acoustic plasmon on the Au (111) surface.
  \emph{Phys. Rev. Lett.} \textbf{2010}, \emph{105}, 016801\relax
\mciteBstWouldAddEndPuncttrue
\mciteSetBstMidEndSepPunct{\mcitedefaultmidpunct}
{\mcitedefaultendpunct}{\mcitedefaultseppunct}\relax
\EndOfBibitem
\end{mcitethebibliography}
\newpage

\begin{figure}[hbt]
\renewcommand\thefigure{}
\centering
\resizebox{0.5\hsize}{!}{\includegraphics*{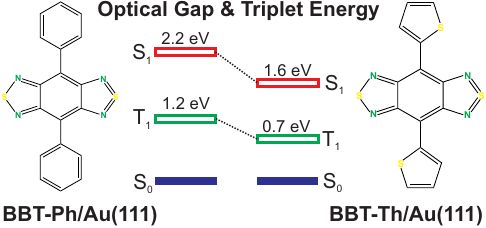}}
\caption{TOC Graphic}
\end{figure}

\end{document}